\pdfoutput=1
\documentclass[twocolumn]{aastex63}
\usepackage{xcolor}

\usepackage{amsmath}
\usepackage{enumitem}
\usepackage{hyperref}
\hypersetup{colorlinks,linkcolor={blue},citecolor={blue},urlcolor={blue}} 
\usepackage{booktabs}

\let\tablenum\relax
\usepackage{siunitx}

\def\msun{\textit{$\rm M_\odot$}}

\def\lbol{\textit{$L_{\rm bol}$}}
\def\edd{\textit{$\lambda_{\rm Edd}$}}
\def\loglbol{\textit{${\rm log}\,L_{\rm bol}$}}

\def\<{$<$}
\def\>{$>$}
\def\g{$g$}
\def\r{$r$}
\def\i{$i$}
\def\z{$z$}
\def\y{$y$}
\def\j{$J$}

\def\ang{$\overset{\circ}{\rm A}$}

\def\hst{{HST}}

\def\hscpipe{{\tt{hscpipe}}}
\def\lens{{\textsc{Lenstronomy}}}

\def\sid{S$\acute{e}$rsic index}
\def\sis{S$\acute{e}$rsic indices}
\def\ss{S$\acute{e}$rsic}

\def\sid{S\'ersic index}
\def\sis{S\'ersic indices}
\def\ss{S\'ersic}

\def\logm{\textit{${\rm log}\,M_*$}}
\def\sigmae{\textit{$\Sigma_{e}$}}

\def\logmbh{\textit{${\rm log}\,M_{\rm BH}$}}
\def\logedd{\textit{${\rm log}\,\lambda_{\rm Edd}$}}

\def\m{\textit{$M_*$}}
\def\mbh{\textit{$M_{\rm BH}$}}
\def\n{\textit{$n$}}
\def\e{\textit{$\epsilon$}}
\def\re{\textit{$r_{\rm e}$}}
\def\qmag{\textit{$Q_{\rm mag}$}}

\def\fr{\textit{$f_{\rm gal}$}}
\def\phot{{\tt{photutils}}}

\shorttitle{Size of Quasar Host Galaxies}
\shortauthors{Li et al.}

\graphicspath{{./}{figures/}}

\begin{document}

\title{The Sizes of Quasar Host Galaxies with the Hyper Suprime-Cam Subaru Strategic Program}

\correspondingauthor{Junyao Li}
\email{lijunyao@mail.ustc.edu.cn}

\author{Junyao Li}
\affiliation{CAS Key Laboratory for Research in Galaxies and Cosmology, Department of Astronomy, University of Science and Technology of China, Hefei 230026, China}
\affiliation{Kavli Institute for the Physics and Mathematics of the Universe, The University of Tokyo, Kashiwa, Japan 277-8583 (Kavli IPMU, WPI)}
\affiliation{School of Astronomy and Space Science, University of Science and Technology of China, Hefei 230026, China}

\author{John D. Silverman}
\affiliation{Kavli Institute for the Physics and Mathematics of the Universe, The University of Tokyo, Kashiwa, Japan 277-8583 (Kavli IPMU, WPI)}
\affiliation{Department of Astronomy, School of Science, The University of Tokyo, 7-3-1 Hongo, Bunkyo, Tokyo 113-0033, Japan}

\author{Xuheng Ding}
\affiliation{Kavli Institute for the Physics and Mathematics of the Universe, The University of Tokyo, Kashiwa, Japan 277-8583 (Kavli IPMU, WPI)}

\author{Michael A. Strauss}
\affiliation{Department of Astrophysical Sciences, Princeton University, 4 Ivy Lane, Princeton, NJ 08544, USA}

\author{Andy Goulding}
\affiliation{Department of Astrophysical Sciences, Princeton University, 4 Ivy Lane, Princeton, NJ 08544, USA}

\author{Simon Birrer}
\affiliation{Kavli Institute for Particle Astrophysics and Cosmology and Department of Physics, Stanford University, Stanford, CA 94305, USA}

\author{Hassen M. Yesuf}
\affiliation{Kavli Institute for the Physics and Mathematics of the Universe (WPI), The University of Tokyo, Kashiwa, Chiba 277-8583, Japan}
\affiliation{Kavli Institute for Astronomy and Astrophysics, Peking University, Beijing 100871, China}

\author{Yongquan Xue}
\affiliation{CAS Key Laboratory for Research in Galaxies and Cosmology, Department of Astronomy, University of Science and Technology of China, Hefei 230026, China}
\affiliation{School of Astronomy and Space Science, University of Science and Technology of China, Hefei 230026, China}

\author{Lalitwadee Kawinwanichakij}
\affiliation{Kavli Institute for the Physics and Mathematics of the Universe, The University of Tokyo, Kashiwa, Japan 277-8583 (Kavli IPMU, WPI)}

\author{Yoshiki Matsuoka}
\affiliation{Research Center for Space and Cosmic Evolution, Ehime University, 2-5 Bunkyo-cho, Matsuyama, Ehime 790-8577, Japan}

\author{Yoshiki Toba}
\affiliation{Department of Astronomy, Kyoto University, Kitashirakawa-Oiwake-cho, Sakyo-ku, Kyoto 606-8502, Japan}
\affiliation{Academia Sinica Institute of Astronomy and Astrophysics, 11F of Astronomy-Mathematics Building, AS/NTU, No.1, Section 4, Roosevelt Road, Taipei 10617, Taiwan}
\affiliation{Research Center for Space and Cosmic Evolution, Ehime University, 2-5 Bunkyo-cho, Matsuyama, Ehime 790-8577, Japan}

\author{Tohru Nagao}
\affiliation{Research Center for Space and Cosmic Evolution, Ehime University, 2-5 Bunkyo-cho, Matsuyama, Ehime 790-8577, Japan}

\author{Malte Schramm}
\affiliation{Graduate School of Science and Engineering, Saitama University, 255 Shimo-Okubo, Sakura-ku, Saitama City, Saitama 338-8570, Japan}

\author{Kohei Inayoshi}
\affiliation{Kavli Institute for Astronomy and Astrophysics, Peking University, Beijing 100871, China}

\begin{abstract}
The relationship between quasars and their host galaxies provides clues on how supermassive black holes (SMBHs) and massive galaxies are jointly assembled. To elucidate this connection, we measure the structural and photometric properties of the host galaxies of $\sim$5000 SDSS quasars at $0.2<z<1$ using five-band ($grizy$) optical imaging from the Hyper Suprime-Cam Subaru Strategic Program. An automated analysis tool is used to forward-model the blended emission of the quasar as characterized by the point spread function and the underlying host galaxy as a two-dimensional Sersic profile. In agreement with previous studies, quasars are preferentially hosted by massive star-forming galaxies with disk-like light profiles. Furthermore, we find that the size distribution of quasar hosts is broad at a given stellar mass and the average values exhibit a size -- stellar mass relation as seen with inactive galaxies. In contrast, the sizes of quasar hosts are more compact than inactive star-forming galaxies on average, but not as compact as quiescent galaxies of similar stellar masses. This is true irrespective of quasar properties including bolometric luminosity, Eddington ratio, and black hole mass. These results are consistent with a scenario in which galaxies are concurrently fueling a SMBH and building their stellar bulge from a centrally-concentrated gas reservoir. Alternatively, quasar hosts may be experiencing a compaction process in which stars from the disk and inflowing gas are responsible for growing the bulge. In addition, we confirm that the host galaxies of type-1 quasars have a bias of being closer towards face-on, suggesting that galactic-scale dust can contribute to obscuring the broad-line region.
\end{abstract}

\section{Introduction} \label{sec:intro}

The ubiquity of supermassive black holes (SMBHs) in massive galaxies and correlations between SMBH mass and properties of their host galaxy (e.g., bulge mass, stellar velocity dispersion) suggest that the formation and/or evolution of the two are connected \citep{Ferrarese2000, Gebhardt2000, Kormendy2013}. Feedback from an active galactic nucleus (AGN) via radiative heating, outflows, or jets has been widely considered as providing such a link that regulates or maybe even halts the growth of galaxies (i.e., star formation). However, direct evidence for a causal connection between SMBHs and their hosts remains elusive. 

Since the masses of SMBHs appear to be more strongly correlated with the properties of the bulges of massive galaxies \citep[e.g.,][]{Kormendy2011, Kormendy2013, Yang2019}, it may be that the growth of SMBHs depends on a buildup of the central mass concentration of their host galaxies \citep[e.g.,][]{Ni2019, Ni2020}. This may be due to the presence of cold gas reservoirs in the inner regions of galaxies. However, the mechanism(s) by which gas on kpc scales loses angular momentum and reaches the central SMBH is unknown \citep[e.g.,][]{Bergmann2019}. In general, it is unclear whether structural changes in the galaxy population, known to coincide with changes in star formation \citep[e.g.,][]{Bell2012, Whitaker2017}, influence the growth of the SMBH and subsequent feedback effects.

Much observational effort has been devoted to understanding whether AGN activity is preferentially associated with galaxies of  specific properties.  Such a connection could reveal the role that galaxy structures play in triggering gas inflows. Studies of X-ray and optically-selected AGNs show that the majority reside in disk galaxies, possibly indicating that structures, such as bars or nuclear spirals, play a role in transporting cold gas to the nucleus \cite[e.g.,][]{Gabor2009, Schawinski2011, Kocevski2012, Yue2018, Zhao2019, Li2020}. There is currently much debate as to whether major mergers, capable of efficiently driving gas inwards through tidally induced torques \citep[e.g.,][]{Hopkins2008}, are the driver of luminous AGN activity \citep[e.g.,][]{Villforth2014, Mechtley2016, Villforth2017, Goulding2018, Marian2019, Marian2020}. Even so, the rarity of such events suggests that mergers are not the key mechanism in triggering the general AGN population \citep[e.g.,][]{Silverman2011, Ellison2011, Ellison2019}.  

Given the importance of understanding how AGN activity is triggered, it is necessary to measure the structural properties of galaxies with strong AGN.
Recent observational studies suggest that AGNs are more likely to be found in compact star-forming galaxies \citep[SFGs; e.g.,][]{Rangel2014, Mushotzky2014, Kocevski2017, Chang2017, Silverman2019, Chang2020}, and the average black hole accretion rate appears to be connected with the central surface-mass density of the host \citep{Ni2019, Ni2020}. In particular, the stellar size of AGN hosts at $z\sim1.5$ is found to lie between those of inactive SFGs and quiescent galaxies (QGs) \citep[][]{Silverman2019}, which seems to be inconsistent with the prediction from some feedback models. In addition, AGN hosts show a prevalence for undisturbed disks \citep[e.g.,][]{Schawinski2011,Kocevski2012,Ding2020}, and there appears to be a mass deficit in their bulges as compared to the local $\mbh - M_{\rm *,bulge}$ relation \citep[][]{Jahnke2009,Ding2020}. Taken together, these studies suggest that AGN activity is preferentially triggered by the same processes that are responsible for building the central stellar-mass concentration, such as a compaction process driven by disk instabilities or minor mergers \citep[e.g.,][]{Bournaud2011, Dekel2014, Lapiner2020}. 

Furthermore, several theoretical studies have argued that energetic feedback from AGNs could have an impact on their size growth \citep[e.g.,][]{Fan2008, Fan2010, Ishibashi2013, Ishibashi2014}. For example, the galaxy size could be affected by mass loss driven by  quasar feedback and the subsequent rearrangement of stellar orbits in response to the change in the depth of the galaxy potential well \citep{Fan2008, Fan2010}. Alternatively, the positive feedback induced by AGN outflows may trigger star formation in the outskirts of galaxies, causing an increase in galaxy size \citep{Ishibashi2013, Ishibashi2014}. Numerical simulations have also shown that invoking AGN feedback in the model is crucial to prevent the formation of over-compact bulges, and to replicate the observed size--mass relationship and stellar density profile for QGs \citep[e.g.,][]{Choi2018, Habouzit2019, vanderVlugt2019}.

To better understanding the role that SMBHs play in galaxy evolution, larger samples of AGN are required over a range of redshifts for which we can measure the structural properties of their host galaxies including the luminous quasars where feedback effects are likely strongest. 
However, separating the host galaxy light from a luminous quasar requires images with excellent resolution, an accurate characterization of the point spread function (PSF), and a careful assessment of systematic measurement biases. Most of the aforementioned studies have mainly focused on infrared and X-ray-selected AGNs, either obscured or of moderate luminosity, to mitigate the effect of AGN contamination on galaxy light distributions. There has been far less work presented on the sizes or stellar mass concentrations of the more luminous type-1 quasars, although a number of Hubble Space Telescope (\hst)-based studies have focused on the properties of the hosts of luminous quasars \cite[e.g.,][]{Bahcall1997, Jahnke2004, Sanchez2004, Peng2006}.
While the high angular resolution of \hst~can image AGN hosts at unprecedented quality, its small field of view limits the number of objects that can be imaged, limiting statistical investigations on the dependence of galaxy structure on AGN properties, such as bolometric luminosity, Eddington ratio and black-hole mass, which contain important clues in understanding AGN triggering and feedback \citep{Chen2020}. It is also unclear whether the aforementioned results still hold at low redshifts $z<1$ where galaxies become increasingly bulge-dominated \cite[e.g.,][]{Bruce2014}. 

Wide-field and deep optical imaging on 4--8m class ground-based telescopes provides new opportunities to address the issues outlined above. For instance, the Hyper Suprime-Cam Subaru Strategic Program (HSC-SSP; \citealt{Aihara2018,Miyazaki2018}) is an ongoing wide-field survey with HSC that will cover 1000 deg$^2$ of the sky in five optical bands ($grizy$). The $5\sigma$ point source depth reaches $\sim$26 mag in $grizy$ with a median seeing of 0.6$''$ in the \i-band over the wide-area component of the HSC-SSP survey, providing unprecedented images for millions of galaxies suitable for morphological and structural analyses \citep{K21}. This combination of depth and high resolution allows us to separate the host galaxy from the bright point source, thus enabling detailed analyses of host-galaxy structures of luminous type-1 quasars with a significantly larger sample than previously achieved.

In a first study, \cite{Ishino2020} analyzed the radial profiles of $\sim1000$ Sloan Digital Sky Survey (SDSS) quasars at $z<1$ with HSC imaging to derive structural and photometric properties for their host galaxies. They found quasar hosts were mainly located in the green valley of the color-magnitude diagram. However, a detailed analysis of the structural properties of the quasar hosts was not presented due to the difficulty of constraining the \ss~parameters from one-dimensional radial profiles and recovering the true galaxy parameters due to systematic measurement biases (see Figure 6 in \citealt{Ishino2020}). Two-dimensional (2D) image analysis, using all the spatial information of an image, is required to better constrain the structural parameters of quasar hosts. Also, simulations are required to correct for systematic biases, particularly for objects with compact or very extended hosts and low host-to-total flux ratios.

In this work, we investigate the structural properties of type-1 quasars selected from the SDSS DR14 quasar catalog \citep{Paris2018} by performing 2D quasar-host decomposition of high-quality HSC images. Here, the principal focus is on the size of their host galaxies, and a comparison with those of inactive galaxies. A companion paper presents the relation between black hole and host galaxy mass and its evolution (Li et al. in preparation). This paper is organized as follows. In Section \ref{sec:data} we describe the data sets used in this paper and our initial sample selection. In Section \ref{sec:method} we introduce the image decomposition and spectral energy distribution (SED) fitting methods, which we use to derive structural and stellar population properties for quasar host galaxies. In Section \ref{sec:simu} we describe the image simulations that we performed to calibrate and assess our structural measurements and SED fitting results. The final sample selection is presented in Section \ref{sec:final_sample}. In Sections \ref{sec:results} and \ref{sec:discussion} we show our results with a focus on the size--stellar mass relation of quasar host galaxies, and discuss their implications in the context of AGN fueling and feedback. In Section~\ref{sec:summary} we summarize our findings. We assume a flat cosmology with $\Omega_\Lambda$ = 0.7, $\Omega_m$ = 0.3, and $H_0 = 70\rm\ km\ s^{-1}\ Mpc^{-1}$. All magnitudes are given in the AB system.

\section{Data and Sample Selection}
\label{sec:data}

\subsection{SDSS DR14 Quasar Sample}
The quasar sample used in this study is drawn from the SDSS DR14 quasar catalog \citep{Paris2018}, which contains 526,357 spectroscopically confirmed quasars that are primarily selected by their optical colors. Additional samples are included based on X-ray, radio, infrared and/or variability detections. The physical properties of the SDSS quasars \citep[][]{Richards2006, Shen2011} used in this study are taken from \cite{Rakshit2019} who performed detailed spectral decomposition and emission-line modeling to measure the bolometric luminosities and black hole masses for the full DR14 sample. The quasar bolometric luminosity (\lbol) is calculated from the monochromatic luminosity using the bolometric correction factors given in \cite{Richards2006}, which are $9.26\times L_{5100}$, $5.15\times L_{3000}$ and $3.81\times L_{1350}$ at $z<0.8$, $0.8<z<1.9$, and $z>1.9$, respectively. 
The black-hole mass of each quasar is estimated using the single-epoch virial estimation method \citep{Vestergaard2002, Vestergaard2006, Shen2011} with the fiducial \mbh~values calculated from the broad ${\rm H}\beta$ emission line at $z<0.8$.
The Eddington ratio is then derived as $\edd = \lbol/L_{\rm Edd}$ where $L_{\rm Edd} = 1.26\times10^{38}\,\mbh$.

\begin{figure}
\centering
\includegraphics[width=\linewidth]{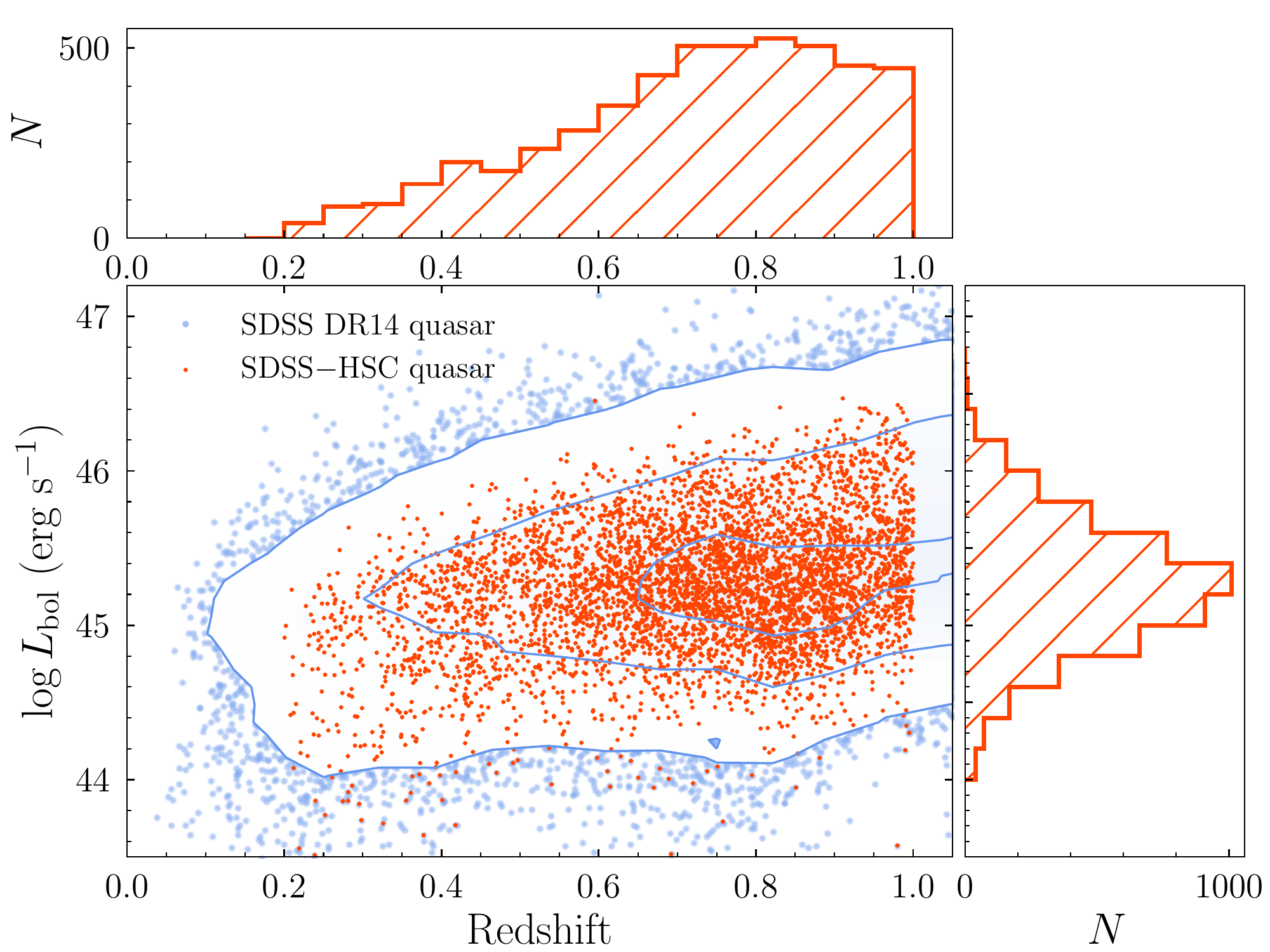}
\caption{Bolometric luminosity as a function of redshift for SDSS quasars. The red points and histograms indicate the 4887 SDSS-HSC quasars used in this work. The blue contours and points show the entire SDSS DR14 quasar population at $z<1$.}
\label{fig:sample}
\end{figure}

\subsection{HSC Imaging Data}

We use the Second Data Release (DR2) of the HSC-SSP \citep{Aihara2019} to carry out structural analyses for SDSS quasars that fall within the wide-area region. A parent quasar sample is constructed by extracting clean HSC objects from the database\footnote{\url{https://hsc-release.mtk.nao.ac.jp/doc/}} that meet the following criteria in all five optical bands:

\begin{enumerate}
    \item isprimary = True
    \item nchild = 0 
    \item pixelflags\_edge = False
    \item pixelflags\_bad = False
    \item pixelflags\_crcenter = False
    \item pixelflags\_bright\_object = False
    \item pixelflags\_bright\_objectcenter = False
    \item specz\_flag\_sdss\_dr14 = True
\end{enumerate}

These selection criteria ensure the following: (1,2) the selected sources are not duplicated or blended, (3-7) their images are away from the edge of the CCD, do not contain bad pixels, and are not influenced by cosmic rays at their cores or near bright objects, and (8) they are included in the SDSS DR14 data release \citep{Abolfathi2018}. 
We then cross-matched the HSC source positions with the SDSS DR14 quasar catalog \citep{Paris2018} using a 1\arcsec~matching radius to identify a sample of 4965 spectroscopically-confirmed quasars at $0.2<z<1.0$ with HSC imaging. The matched quasars include 189 with a saturated central core ({{\tt pixelflags\_saturatedcenter = True}}) in at least one of the five HSC bands.
We further excluded 78 sources through visual inspection because they are affected by large and bright nearby companions in projection (21 objects), cosmic rays not flagged appropriately in the database (3 objects), close to the CCD edge but not flagged appropriately in the database (4 objects), the close companians and the target quasar are misidentified as one object (25 objects), or that the entire region within the FWHM of the PSF are labeled as saturated (25 objects). This leaves 4887 objects (hereafter the full sample).
We impose a lower limit on the redshift at $z=0.2$ because significant substructures (e.g., spiral arms) seen in low-redshift galaxies cannot be well described by a single \ss~profile which we adopt as our fiducial model to fit the host galaxy (see Section \ref{subsec:decomp_method}). The upper redshift bound ($z=1$) is set because surface brightness dimming results in lower values of the host-to-total flux ratio that make the host galaxy difficult to detect \citep{Ishino2020}. The distribution of redshift and bolometric luminosity are shown in Figure~\ref{fig:sample} for the matched SDSS-HSC quasar sample relative to the overall SDSS quasar population.

 The co-added HSC images in $grizy$ bands \citep{Kawanomoto2018}, together with the variance images, mask images and PSF models for each source are generated using the HSC software pipeline \hscpipe~v6.7 \citep{Bosch2018}. In general, candidate stars ($\sim70$) are selected to characterize the PSF at a given position on each CCD using a second-order polynomial \citep{Aihara2018, Coulton2018, Carlsten2018}. Each co-added image has a global background level  subtracted based on a 2D Chebyshev polynomial fit to a binned image with a typical box size of $256\times256$ pixels \citep{Aihara2019}. To account for background residuals and variations on small scales that are not captured by the standard HSC background subtraction, an additional local background subtraction is performed using the $\tt{SExtractorBackground}$ algorithm built in the Python package \phot~with the box size being chosen on the basis of the typical source size in the image ($\sim 50\times50$ pixels). 
 
 \subsection{Comparison with Inactive Galaxies}
 \label{subsec:control}
 To put quasar hosts in the general context of galaxy evolution, we used galaxy samples from the literature to make direct comparisons between the properties of active and inactive galaxies. The first comparison sample was drawn from \cite{K21} (hereafter K21), which presents structure measurements for $\sim1.5$ million galaxies from the Second Data Release of the HSC survey. Structural parameters of galaxies in K21 are measured by fitting model 2D \ss~profiles to HSC \i-band images with the same tool \lens\footnote{\url{https://github.com/sibirrer/lenstronomy}} \citep{Birrer2015, Birrer2018} as implemented in this study (see Section \ref{subsec:decomp_method}). Photometric redshifts, stellar masses and rest-frame colors are derived using MIZUKI \citep{Tanaka2015} and EAZY \citep{Brammer2008} with HSC photometry, which assume the same SED fitting models as adopted in this work (see Section \ref{subsec:sed}).
 K21 performed detailed image simulations and statistical modeling to account for systematic measurement biases, catastrophic photometric redshift failures, and cross-contamination between star-forming and quiescent galaxies. This catalog presented an ideal comparison sample with a large number of massive galaxies in which quasars usually reside.
 
A second comparison sample was taken from \cite{Mowla2019} (hereafter M19), which combines 33,879 galaxies from the five CANDELS fields and 918 massive galaxies ($\logm>11.3$) from the COSMOS-DASH survey. The CANDELS and COSMOS-DASH catalogs include the following: spectroscopic redshifts where available \citep{Brammer2012, Momcheva2016}, photometric redshifts estimated by EAZY, information on the stellar population (e.g., $M_*$, SFR, rest-frame colors) estimated by FAST \citep{Kriek2009, Skelton2014} while assuming the same SED fitting models as our study, and galaxy structural properties provided by \cite{vanderwel2014} and M19. Structural parameters are determined by fitting model 2D \ss~profiles to HST/WFC3 images in the F125W and F160W filters using GALFIT \citep{Peng2002}. The wealth of multi-wavelength photometry in the CANDELS and COSMOS fields as well as high resolution HST images allow accurate measurements of photometric redshifts, galaxy structural parameters and stellar population properties that are used here to evaluate the reliability of our decomposition results (see Section \ref{subsec:evaluate}).
 
\begin{figure*}
\centering
\includegraphics[width=0.8\linewidth]{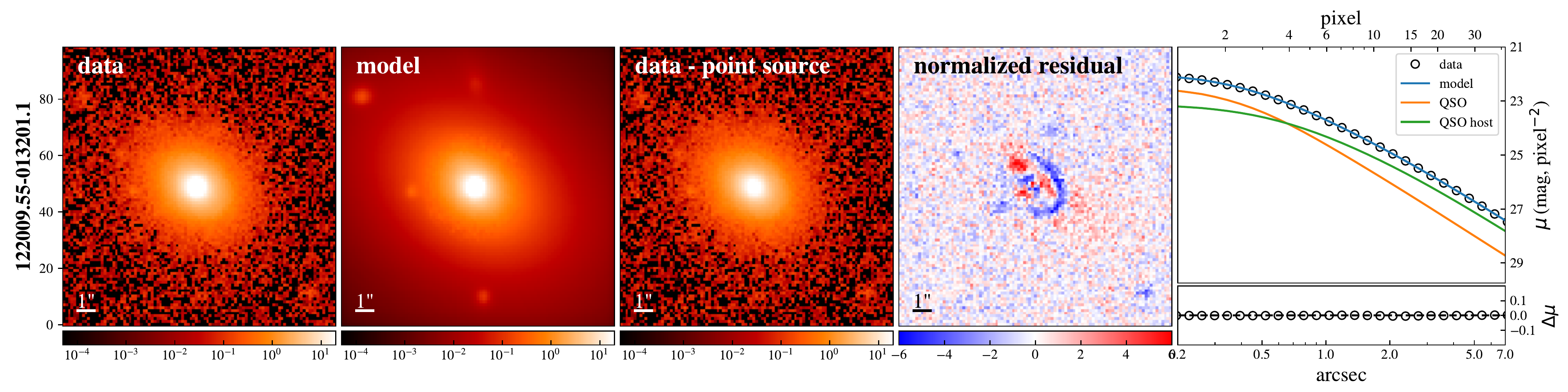}
\includegraphics[width=0.8\linewidth]{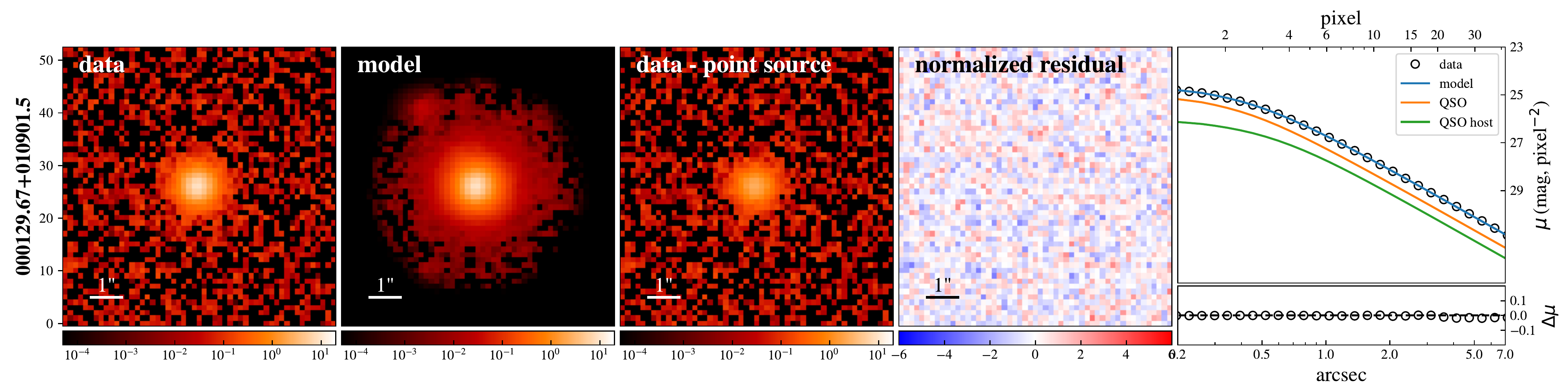}
\includegraphics[width=0.8\linewidth]{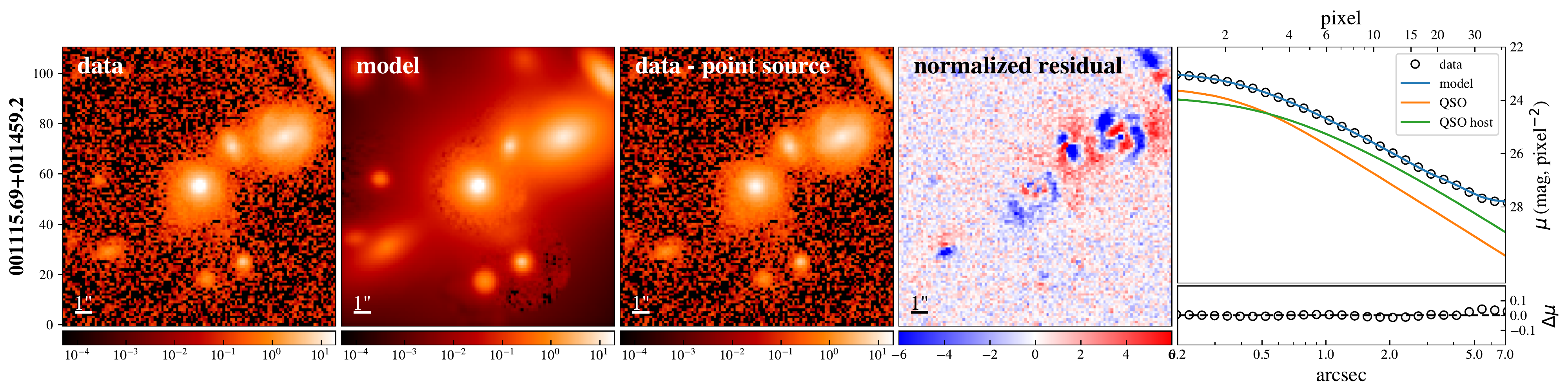}
\includegraphics[width=0.8\linewidth]{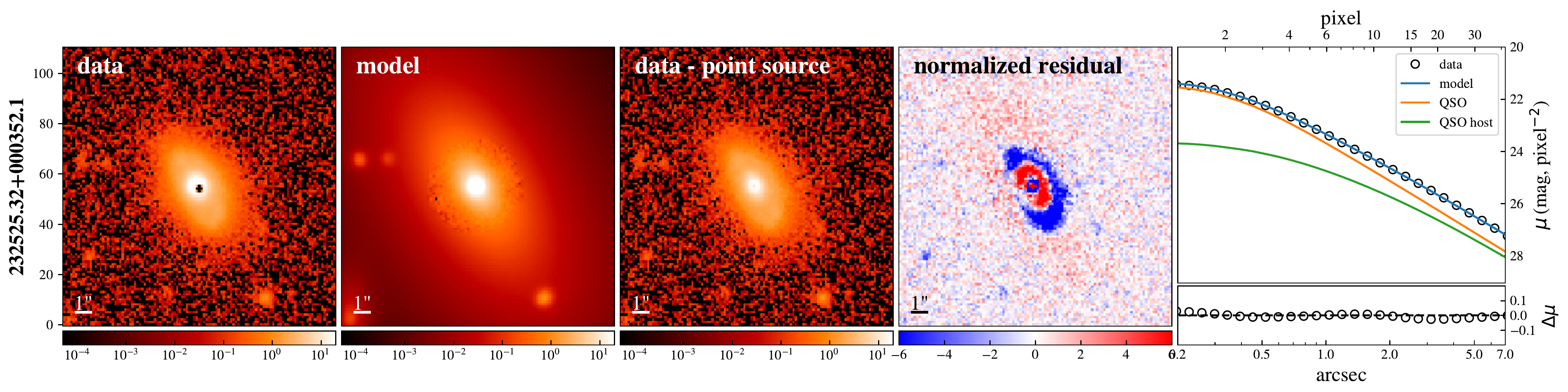}

\caption{Representative examples of quasar-host galaxy decomposition using HSC \i-band images. The panels from top to bottom display an isolated quasar at $z=0.29$ (first row) and $z=0.96$ (second row), a $z=0.58$ quasar in a crowded field (third row) and a quasar at $z=0.34$ (fourth row) for which saturated pixels (central black dot in the left-most panel) are masked in the fitting routine. The panels from left to right are as follows: (1) observed HSC \i-band image; (2) best-fit point source plus host galaxy model; (3) data minus the point source model (the pure galaxy image); (4) fitting residual divided by the variance map; (5) one-dimensional surface brightness profile (top) and the corresponding residual (bottom). For the images, the color scale is in counts s$^{-1}$.}
\label{fig:example}
\end{figure*}

\section{Methods}
\label{sec:method}

\subsection{2D Image Decomposition and Fitting}
\label{subsec:decomp_method}

We implement an automated fitting routine using the image modeling tool \lens~to derive galaxy structural properties from HSC images. \lens~is a multipurpose package initially designed for gravitational lensing analysis through forward modeling. However, its high-level of flexibility enables us to decompose images into quasar and host contributions based on 2D profile fitting. 
\lens\ uses Particle Swarm Optimization \citep{Kennedy1995} for $\chi^2$ minimization, to reduce the likelihood of getting trapped in a local minimum when searching the full parameter space, and Markov chain Monte Carlo (MCMC) 
for Bayesian parameter inference (emcee; \citealt{Mackey2013}). 

The inputs to \lens~include the cut-out image, noise map, and empirical model PSF image, from which  seeing-corrected quantities are derived. The size of the cut-out image is chosen automatically to be between $41\times41$ pixels ($7\arcsec\times7\arcsec$) and $131\times131$ pixels ($22\arcsec\times22\arcsec$) to ensure that the flux from the target quasar and all close companions are included. Our sample contains 189 quasars with a saturated central core, for which we mask the saturated pixels in the fitting. For quasars with companion galaxies, we simultaneously fit, using a single \ss~model for each, nearby objects within each cutout that are brighter than 25 magnitude as indicated by the HSC source catalog.

We then perform 2D profile fitting to the background-subtracted cut-out images to derive structural properties of quasar host galaxies. A 2D \ss~profile \citep{Sersic1963} convolved with the PSF is used to model the galaxy component. Its 1D projection is parameterized as 
\begin{equation}
I(r) = I_e\, {\rm exp}\left(-b_n \left[\left(\frac{r}{r_e}\right)^{1/n} - 1\right]\right),
\end{equation}
where \re~is the galaxy half-light radius along the major axis, adopted as an indicator of galaxy size, and \n~is the \sid~with the constant $b_n$ being uniquely determined for a given \n. The normalization of the profile is determined by $I_e$ which represents the flux intensity at \re. Two additional parameters are the ellipticity \e~and position angle $\phi$ that describe the galaxy shape and orientation. We limit the parameters of the \ss~model as follows: $0.1\arcsec < \re < 5\arcsec$, $0.3 < n < 7.0$ and $0.0 < \epsilon < 0.85$. We remark that our fitting routine only models the smooth component of a galaxy, and does not attempt to model substructures such as spiral arms, bars and tidal tails.

The point source component is fit using a PSF model. The position of the point source can differ from the galaxy center by up to $0.3\arcsec$. While \lens~can take an input 2D error image of the PSF into account, the median error on the fractional PSF model size in the \i-band is less than 0.4 percent \citep{Mandelbaum2018}, and thus we consider a fixed PSF per source in our modeling.
However, it is possible that the PSF model for an individual quasar may not be perfect due to the color difference between the quasar and the stars that are used to construct the PSF. 

The \i-band images are preferentially taken when the seeing is better than 0.7\arcsec~to satisfy requirements on galaxy shape measurements for weak lensing analysis thus the median seeing is 0.6\arcsec~\citep{Mandelbaum2018, Aihara2019}. We thus first fit the $i$-band image for each quasar since the superb quality of these images allows a more robust decomposition of an unresolved point source and model host galaxy. We then fit the \g-, \r-, \z-~and \y-band images by fixing the structural parameters to the \i-band results. The positions of each component are set as free parameters. The derived quasar-subtracted host-galaxy fluxes (after performing the calibrations in Section \ref{sec:simu}) are used to construct host SEDs and derive the properties of the stellar population (see Section \ref{subsec:sed}). The \ss~parameters derived from the \i-band images are used in the following galaxy structural analyses. Note that the \i-band images probe different rest-frame wavelengths, and it is well known that the galaxy structural properties are wavelength-dependent \citep[e.g.,][]{vanderwel2014, Vulcani2014, Kennedy2016}. Therefore, we follow \cite{vanderwel2014} to correct our sizes (after performing the calibrations in Section \ref{sec:simu}) to a common rest-frame wavelength of 5000~\ang. The details are given in Appendix~\ref{appendix:a}.

In Figure \ref{fig:example} we show four representative examples of the decomposition results including the original \i-band science image, the total model image convolved with the PSF, the point source-subtracted galaxy image, the normalized residual map, and the one-dimensional surface brightness profile shown from left to right, respectively. The four images from top to bottom are for an isolated quasar at $z=0.29$, an isolated quasar at $z=0.96$, a quasar in a crowded field at $z=0.58$ and a saturated quasar at $z=0.34$ for which the saturated central pixels are masked in the fitting. 

The 1$\sigma$ confidence interval of the best-fit structural parameters and fluxes of each component can be inferred through a MCMC sampling procedure. However, the derived $1\sigma$ uncertainties are extremely small and likely to be underestimated, given that we adopt a fixed PSF model for each quasar and fix the parameters to the \i-band results in the fitting.
In order to quantify the goodness of fit to the quasar itself, we define a quantity $\Delta {\rm SB}$ as the rms value of $\Delta \mu$ within $R_{\rm in}$, where $R_{\rm in}=2\,r_e$ and $\Delta \mu$ is the residual of the one-dimensional surface brightness profile (see the fifth column in Figure \ref{fig:example}). For cases where $\re < 0.2''$, $R_{\rm in}$ is set to $0.4''$.

We note that due to model degeneracy, the code, in some cases, treats an unresolved source as a bright compact galaxy with a much fainter point source, thus the host-to-total flux ratio (\fr) is close to one. To determine whether a host is actually detected, we also fit our sample with the PSF model only, and compare with the model which includes the host component using the Bayesian Information Criterion (BIC; \citealt{schwarz1978}). The BIC value for a given model is defined as ${\rm BIC} = k\, {\rm ln}(n) - 2\, {\rm ln} (\hat{L})$, where $k$ is the number of model parameters, $n$ is the number of effective data points (i.e., excluding masked pixels) and $\hat{L}$ is the maximized value of the likelihood function.
We define $\Delta \rm BIC \equiv BIC_{\rm psf+host} - BIC_{psf}$ and use $\Delta \rm BIC>10$ \citep{Kass1995} as a threshold value to determine whether a host component is strongly required.
The values of $\Delta \rm SB$ and $\Delta \rm BIC$ will be used to select our final quasar sample in Section \ref{sec:final_sample}.

\subsection{SED Fitting}
\label{subsec:sed}

\begin{figure}
\centering
\includegraphics[width=\linewidth]{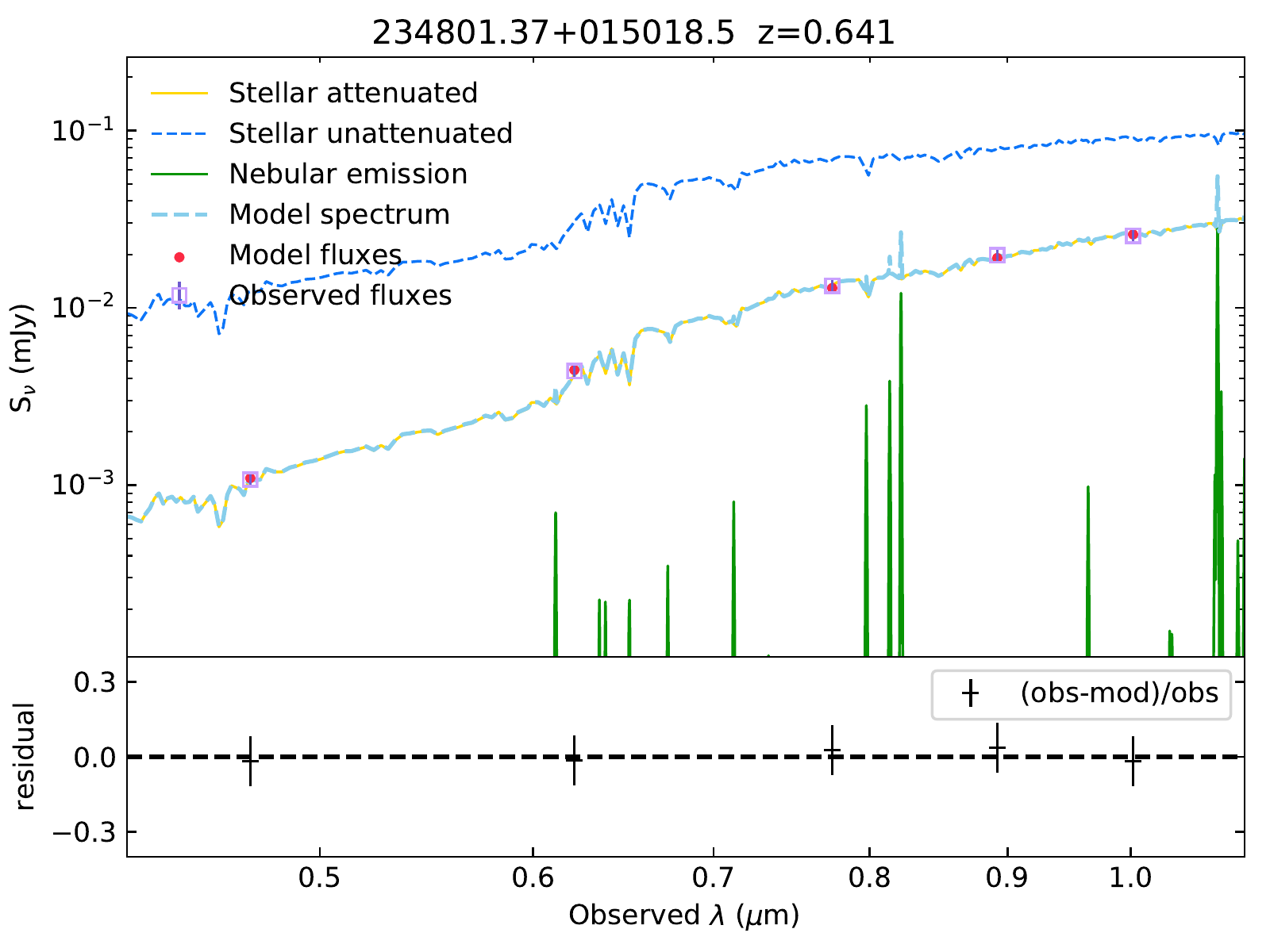}
\caption{Example SED of a quasar host galaxy constructed from the decomposed optical emission of SDSS J234801.37+015018.5 ($z=0.641$) in five HSC bands with the best-fit SED of a massive ($\logm/M_\odot = 10.8$, $E(B-V)=$ 0.39) galaxy overlaid. The bottom panel displays the residuals between the data and model.}
\label{fig:sed}
\end{figure}

We determine the best-fit SED for each quasar host galaxy using the decomposed galaxy fluxes, free of quasar emission, in each of the $grizy$ bands. The flux is calibrated in each band using the simulations presented in Section \ref{sec:simu} to account for measurement biases. Since the errors derived through MCMC sampling by \lens~are likely underestimated, we assign a 10 percent error to the galaxy flux, which is the typical fractional dispersion of the input-to-output flux ratio as determined by our simulations (Section \ref{sec:simu}). We then use the SED fitting code CIGALE \citep{Boquien2019} to derive stellar mass $M_*$ and rest-frame colors of the quasar host galaxies. The model SEDs incorporate a delayed star formation history (SFH), stellar population synthesis models \citep{Bruzual2003}, a \cite{Chabrier2003} initial mass function, a \cite{Calzetti2000} attenuation law, and a nebular emission model. An example of the model fit to the decomposed five-band photometry is shown in Figure \ref{fig:sed}.

\begin{figure}
\centering
\includegraphics[width=\linewidth]{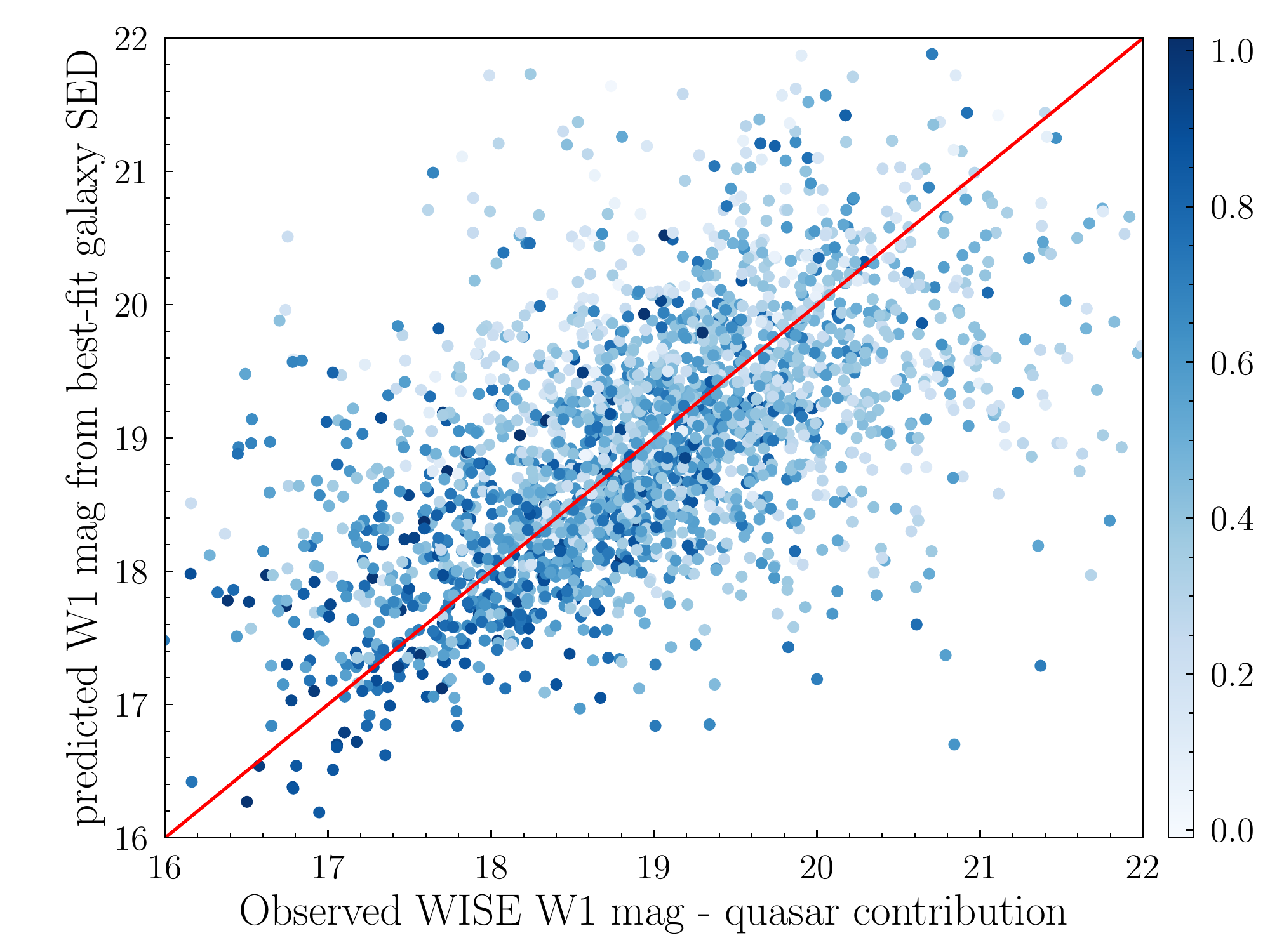}
\caption{Comparison of the host galaxy magnitudes in the WISE W1 band extrapolated from the best-fit SED to the observed W1 magnitudes after subtracting the quasar contribution. Each data point is color coded by the \i-band host-to-total flux ratio.}
\label{fig:wise}
\end{figure}

We independently test the accuracy of our SED fitting results by comparing the host-galaxy fluxes at MIR wavelengths, based on an extrapolation from the best-fit SED, to those observed by the Wide-field Infrared Survey Explorer (WISE; \citealt{Wright2010}). All of our quasars are detected at mid-infrared (MIR) wavelengths by WISE. We particularly use the W1 band at 3.4~$\rm \mu m$ as the observed WISE flux is not totally dominated by the quasar emission. We use the type-1 quasar template of  \cite{Lyu2017} and our decomposed quasar magnitude in the \i-band to estimate the quasar contribution to the WISE W1 band. The fraction of the W1 emission attributed to the quasar spans a broad range with a median contribution of 48\%. We then subtract the expected quasar contribution from the observed W1 magnitude, and compare with the pure galaxy flux derived from the best-fit SED. As shown in Figure \ref{fig:wise}, the two values are in good agreement on average, suggesting that we do not systematically overestimate or underestimate the host fluxes. However, there is a substantial amount of scatter due to the uncertainties of constraining rest-frame IR fluxes which are not covered by our HSC data, as well as the fact that we only subtract an average quasar SED in our estimation.

The rest-frame $U-V$ color is then derived from the best-fit SED. We use the $U-V$ vs. $M_*$ diagram instead of the commonly used $UVJ$ diagram \citep{Wuyts2007, Williams2009} to separate quasar hosts into star-forming and quiescent galaxies: the rest-frame $J$-band fluxes are not well constrained in our fitting while the stellar masses are  robust (see Section \ref{subsec:evaluate}). We develop our $UV-M_*$ criteria based on the classification result from the $UVJ$ diagram. We first label each CANDELS galaxy as star-forming or quiescent based on their $UVJ$ colors, and then use the Support Vector Machine algorithm to derive the 
boundary on the $U-V$ vs. $M_*$ plane that most accurately classifies the galaxy types.
The boundary line is parameterized as $U-V = 0.16 \times \logm + 0.16$, which will be used to 
separate quasar hosts into star-forming and quiescent galaxies in Sections \ref{subsec:evaluate} and \ref{subsec:sf_prop}. A more detailed discussion of the reliability of our SED fitting results and the classification of galaxy types will be presented in Section \ref{subsec:evaluate}.

We provide a full catalog of our decomposition and SED fitting results of 4887 quasars including the calibrations described in Section \ref{sec:simu}. Table \ref{table:catalog} gives a description of the columns in the catalog.

\section{Calibration and assessment of best-fit parameters}
\label{sec:simu}

Structural measurements of galaxies are prone to significant biases (e.g., an underestimation of size for large galaxies) due to PSF blurring, finite signal-to-noise ratio and surface brightness dimming \citep[e.g.,][]{Carollo2013, Tarsitano2018}. Measurements based on images that are taken from ground-based telescopes may be more severely affected by the aforementioned issues, and the presence of an additional unresolved point source component which can be degenerate with the centrally concentrated bulge makes the measurements even more challenging \citep[e.g.,][]{Bruce2016, Ishino2020}. To correct for inherent biases, we use image simulations for which we know precisely the intrinsic properties of simulated galaxies to assess and calibrate our measurements. We take two approaches: in the first, we use model galaxies inserted into empty regions of real HSC images, and in the second we add an unresolved quasar using HSC model PSFs to real HSC images of galaxies in the CANDELS field.

\subsection{Using Model Galaxy Images}
\label{subsec:simu}

\begin{figure*}
\centering
\includegraphics[width=\linewidth]{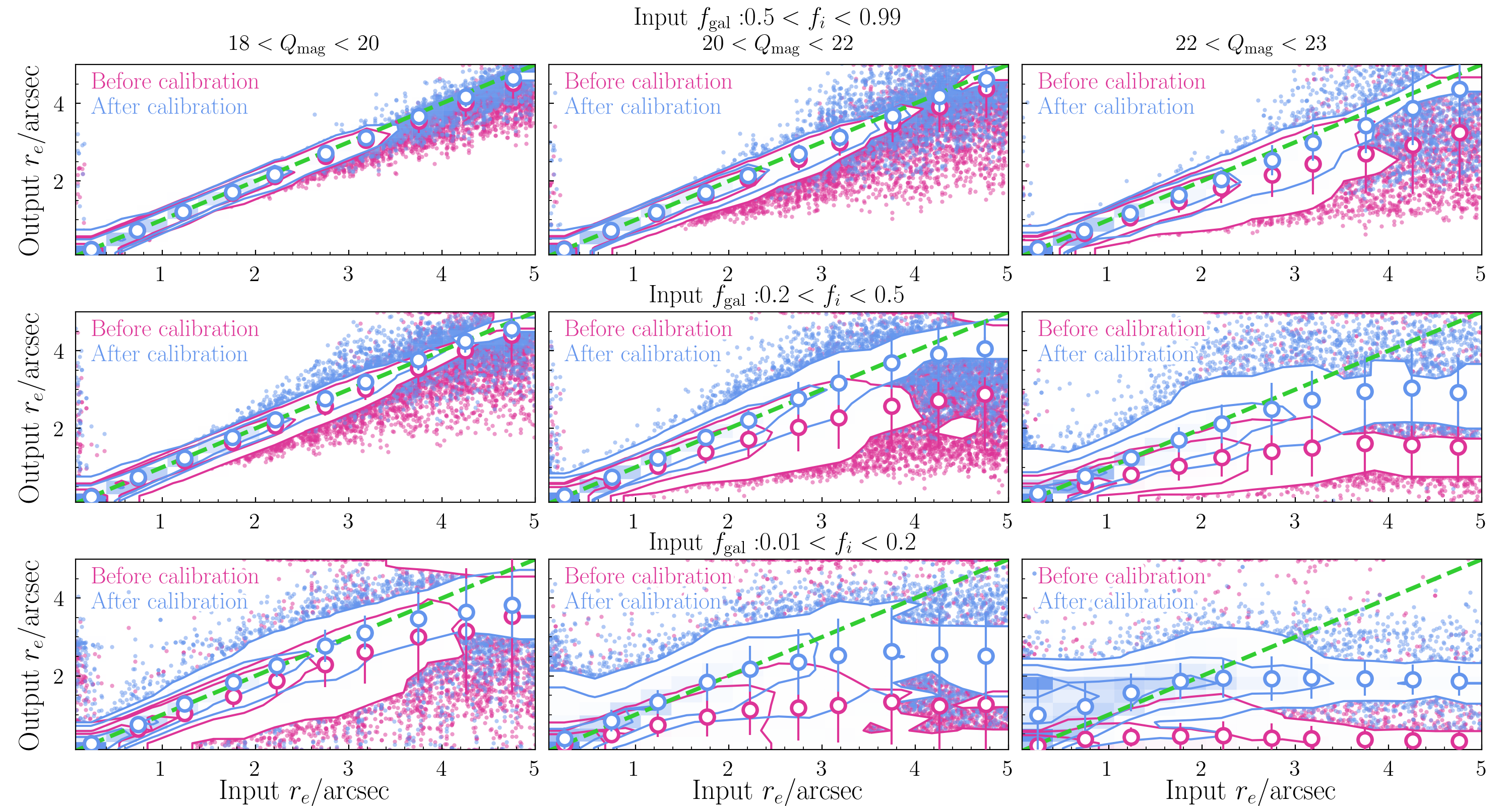}
\caption{Simulation results (I) using models for both the quasars and their host galaxies. Comparisons of input and output \i-band size (\re) for simulated galaxies in three different input quasar magnitude bins (columns) under the median seeing condition (0.6\arcsec). The range of input host-to-total flux ratio is 0.01 to 0.99, split as indicated in each row. The calibrated (uncalibrated) results for individual sources are shown as blue (magenta) contours and points. The uncalibrated and calibrated median output values as a function of input values are shown as magenta and blue open circles, respectively. Error bars are given as 1.4825 $\times$ median absolute deviation. The calibration procedure is able to effectively reduce the systematic measurement biases, particularly with respect to the underestimation of size for large galaxies.}
\label{fig:input_output}
\end{figure*}

We used \lens~to simulate a series of \i-band quasar images represented by a PSF on a magnitude grid of \qmag~= [18, 19, 20, 21, 22, 23, 24]. Three PSF models representing medium (0.60\arcsec), better (0.45\arcsec), and worse (0.75\arcsec) HSC $i$-band seeing conditions are used in the simulations. We randomly sampled 30 points within each magnitude grid. For each model quasar, a host-galaxy image was generated based on a 2D \ss~profile that sampled a grid in parameter space (\fr, \re, \n, \e; Section \ref{subsec:decomp_method}). The grids span the following intervals: $\fr = [0.01, 0.1 - 0.9~{\rm (in~steps~of~0.1)}, 0.99]$, $r_e = [0.07, 0.25, 0.55, 0.9, 1.3, 1.8, 2.3, 3.2, 5.0]\arcsec$, $n = [0.3, 0.6, 0.9, 1.3, 1.7, 2.1, 2.7, 3.5, 4.5, 7.0]$ and $\epsilon = [0.0, 0.15, 0.3, 0.5, 0.8]$. For each grid cell, we randomly sampled 30 parameters to create model galaxy images with each convolved with the appropriate PSF model and added to the quasar image along with poisson noise. Model images were inserted into empty regions of real HSC \i-band images. The variance map of the HSC image with additional poisson errors was used as the weight map in the fitting procedure. The size and pixel scale of the simulated images are $60\arcsec\times60\arcsec$ and 0.168\arcsec. In total, we generated $\sim 1$~million simulated images.

The decomposition procedure is then run on all \i-band simulated images in the same manner as for the real SDSS quasar images. In Figure~\ref{fig:input_output} we show the comparisons of input and output (fitted) $r_e$ as a function of input \qmag~and \fr~for median seeing. Magenta contours and points illustrate systematic measurement biases in specific regions of parameter space. In each panel, we calculate the median values of the output parameter for each input bin (magenta circles). For sources with the largest host-to-total flux ratios ($0.5<\fr<0.99$), the size can be recovered well. However for the faintest quasars (i.e., $22<\qmag<23$) the fit tends to underestimate \re~for the most extended galaxies ($\re > 3\arcsec$). At moderate host-to-total ratios ($0.2<\fr<0.5$), while the systematic offset and scatter increase, we can still constrain the size well for bright sources ($\qmag<22$), particularly for galaxies with $\re\lesssim2.5\arcsec$. The systematic uncertainties on the measured size are most significant for faint hosts (e.g., $\qmag > 22$~mag and $\fr < 0.2$, corresponding to host galaxies fainter than 23.7 mag) in the sense that we tend to significantly underestimate the size for galaxies of all sizes. The failure to recover structural parameters for galaxies fainter than $\sim23$~mag has been observed in previous HST-based image simulations, due to degeneracies between the faint outer regions of galaxies and the background sky \citep[e.g.,][]{Haussler2007, Carollo2013}. However, we note that 89\% of our final sample (see Section \ref{sec:final_sample}) have quasars brighter than 22~mag, hosts brighter than 23~mag, and sizes smaller than $2\arcsec$ (see Section \ref{sec:results}), thus they rarely fall in the parameter space illustrated by the third column in Figure~\ref{fig:input_output}. Similar comparisons for the \sid, ellipticity and host-galaxy magnitude  are presented in Appendix \ref{appendix:b}.

\subsubsection{Corrections on Structural Measurements}
\label{subsubsec:calibration}

With the suite of intrinsic and fitted parameters for simulated galaxies, we quantified the systematic biases of structural and flux measurements and used these to derive corrections by measuring the differences between fitted and input parameters in a six dimensional space ($\qmag-r_e-n-\fr-\epsilon-{\rm PSF}$) following \cite{Carollo2013}. Specifically, for a given real observed galaxy with a corresponding PSF and a set of fitted parameters (represented by the subscript $t$) $r_{e, t}$, $n_t$, $Q_{{\rm mag},t}$, $f_{{\rm gal}, t}$ and $\epsilon_t$, we identified all simulated objects that have fitted parameters within a grid cell centered on the observed galaxy. The boundaries of the grid cell are chosen as $Q_{{\rm mag},t} \pm 0.5$~mag, $f_{i, t} \pm 0.05$, $\epsilon_t \pm 0.2$, $r_{e, t} \pm 0.3\arcsec$ and $n_t \pm 0.3$. The corrections to be applied for each real galaxy are then determined by taking the median differences between the input model parameters and the fitted values for all model galaxies within a grid cell. 

For cases in which no model galaxies fall within a grid cell corresponding to true galaxy, we perform an iterative procedure that gradually expands the cell size with $\Delta Q_{{\rm mag},t} = 0.1$~mag, $\Delta f_{i, t} = 0.02$, $\Delta \epsilon_t = 0.05$, $\Delta r_{e, t} = 0.1\arcsec$ and $\Delta n_t = 0.1$, to identify model galaxies in larger boxes. The maximum number of allowed iterations is five. Note that there are still 7 sources for which we cannot find analogous model galaxies after implementing this iteration process, thus no corrections can be made and the fits are considered to be unreliable. 
This mainly occurs when the measured galaxy brightness is extremely faint and/or the \sid~hits the boundary of the allowed values.

\begin{figure*}
\centering
\includegraphics[width=\linewidth]{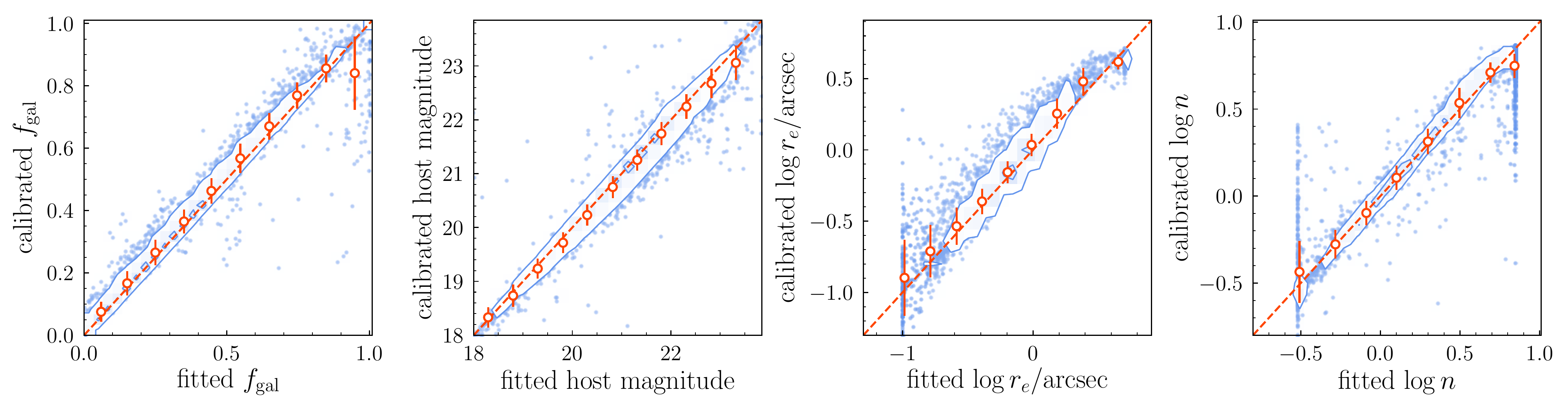}
\caption{Comparison of the $i$-band best-fit model parameters (no correction applied) to the calibrated values (see Section~\ref{subsubsec:calibration}) for the 4887 SDSS quasars. The panels from left to right are as follows: (1) host-to-total flux ratio; (2) decomposed host-galaxy magnitude; (3) galaxy size and (4) \sid. The blue contours and points show the individual measurements. The red circles represent the mean calibrated values with $1\sigma$ dispersion.}
\label{fig:cmp}
\end{figure*}

Following \cite{Tarsitano2018}, we quantify the dispersion ($w$) of the cell in the 5D parameter space as the quadratic sum of the variances of the model parameters defined as
\begin{equation}
w = \sqrt{\sum_{i}{\frac{\hat{\sigma}_i^2}{\hat{m}_i^2}}},
\end{equation}
where $i$ = \qmag, \fr, \re, \n, \e, and $\hat{\sigma}_i^2$ and $\hat{m}_i$ are the variance and median values of each model parameters, respectively. A large value of $w$ (e.g., 2.0) indicates that a diversity of simulated galaxies with different input model parameters can scatter into a given cell and display similar observed values, thus the fitting results and the corresponding corrections are unreliable.

We first apply the calibration procedure to simulated galaxies to demonstrate its effectiveness. The calibrated median output sizes are shown in Figure \ref{fig:input_output} as blue circles. Generally speaking, the correction procedure is able to bring the median output sizes back to the one-to-one line, at least for $\re \lesssim 3.0\arcsec$, a parameter space in which the majority ($\sim 95\%$) of our SDSS quasars fall. At low \fr~(e.g., $\fr < 0.2$) and large \re~(e.g., $\re>3.0\arcsec$), the calibrated values still deviate from the one-to-one line, indicating that we are unable to reliably recover the intrinsic sizes even after applying the corrections. This is because for such faint sources, a given output value can be generated by a wide range of input parameters, meaning that there is not a unique
correction value. Even so, the significantly reduced offset after applying corrections indicates that the calibration procedure is still effective. Similar analyses for the \sid, ellipticity and host galaxy magnitude are described in Appendix \ref{appendix:b}.

This calibration routine is then implemented on the observed SDSS quasars. For sources having an associated PSF of size $<0.45\arcsec$, $0.50\arcsec - 0.70\arcsec$ and $>0.70\arcsec$, we use the simulation results for PSF sizes equal to $0.45\arcsec$, $0.60\arcsec$ and $0.75\arcsec$, respectively, to perform the calibrations. On average, ninety simulated quasars are used to calibrate measurements of each SDSS quasar. The comparisons between the fitted and calibrated parameters are shown in Figure \ref{fig:cmp}. 
Overall, the corrections for the host-to-total flux ratio and the host-galaxy magnitude are small, while the sizes tend to be slightly underestimated prior to correction. 
The largest correction occurs when the fitted \sid~hits the lower limit ($n=0.3$) or the upper limit ($n=7.0$) of the values considered. 
In general, it is difficult to distinguish the host component from the quasar and recover the true galaxy parameters in cases where (1) the host is faint,
(2) there are degeneracies between the PSF and a compact galaxy (for large values of $n$ where the light distribution is steep at $r<\re$) or (3) there are considerable errors on the PSF model. For such sources, the calibrated values are largely dominated by the correction term which is uncertain. This is further discussed in Section \ref{subsec:evaluate}.

As a word of caution, since we only simulate the optimal situation, i.e., an isolated quasar with its host galaxy following a perfect single-component \ss~profile, other factors such as contamination from nearby objects, deviations from a single-component \ss~profile and galaxy substructures are not taken into account. Therefore, the corrections we have implemented here may be underestimated. However, in the next section, we carry out a similar exercise using real non-AGN HSC galaxies as the hosts to consider some of the effects not properly addressed using model galaxies.

\subsection{Using Real Galaxy Images}
\label{subsec:evaluate}

\begin{figure*}
\centering
\includegraphics[width=\linewidth]{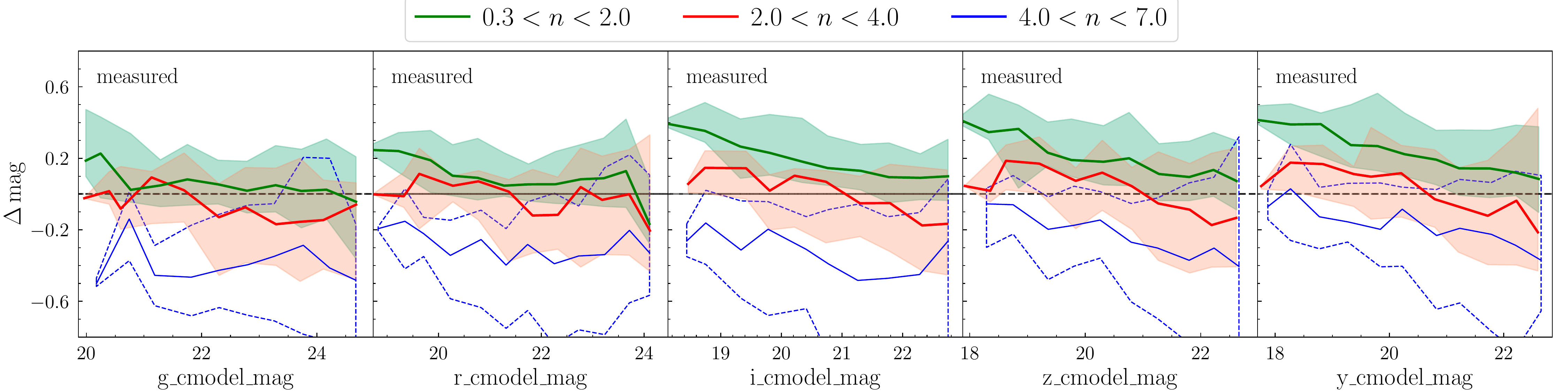}
\includegraphics[width=\linewidth]{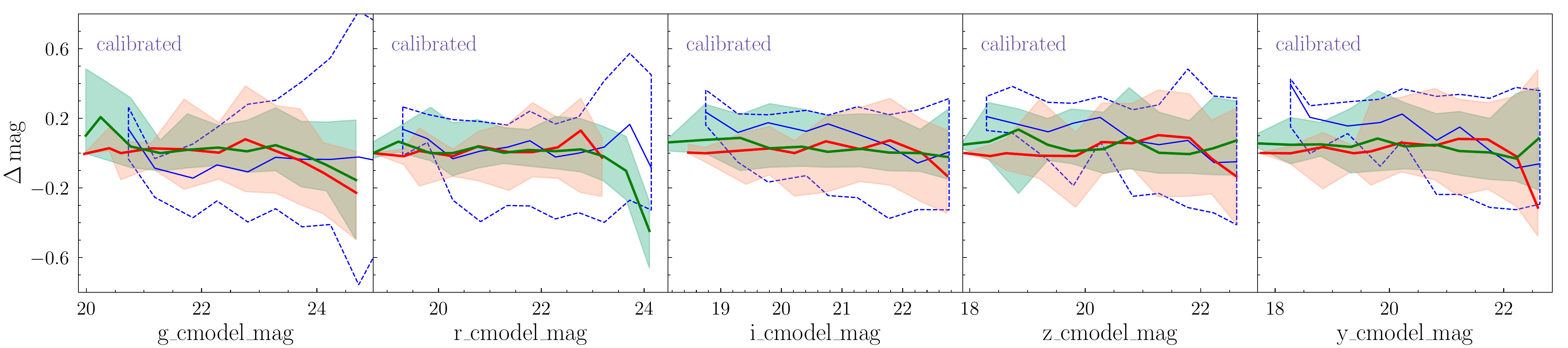}
\caption{Simulation results (II) using model quasars with real inactive CANDELS galaxies as their hosts. Median offset and $1\sigma$ scatter between the decomposed galaxy magnitudes and the true galaxy magnitudes (i.e., the HSC cmodel magnitudes for CANDELS galaxies) as a function of true galaxy magnitudes in three \sid~bins for simulated quasars (see Section \ref{subsec:evaluate}) are plotted. The top row shows the uncorrected measurements. The bottom row shows offsets after applying corrections to the decomposed magnitudes using the method described in Section \ref{subsec:evaluate}. The calibration procedure is able to largely eliminate the systematic offset of our flux measurements, minimizing biases in our stellar mass estimates.
}
\label{fig:cmp_mag}
\end{figure*}

\begin{figure*}
\centering
\includegraphics[width=\linewidth]{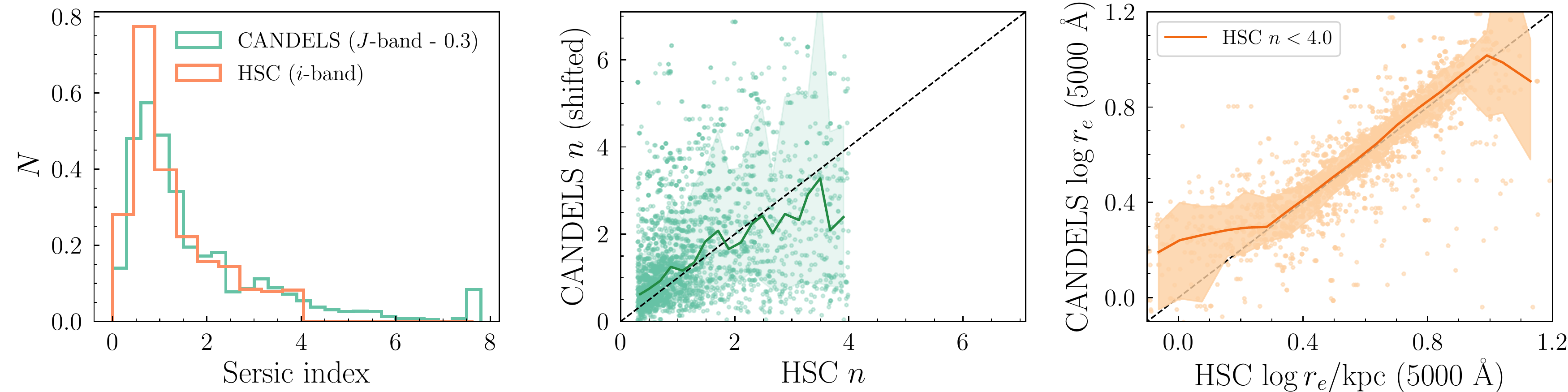}
\includegraphics[width=\linewidth]{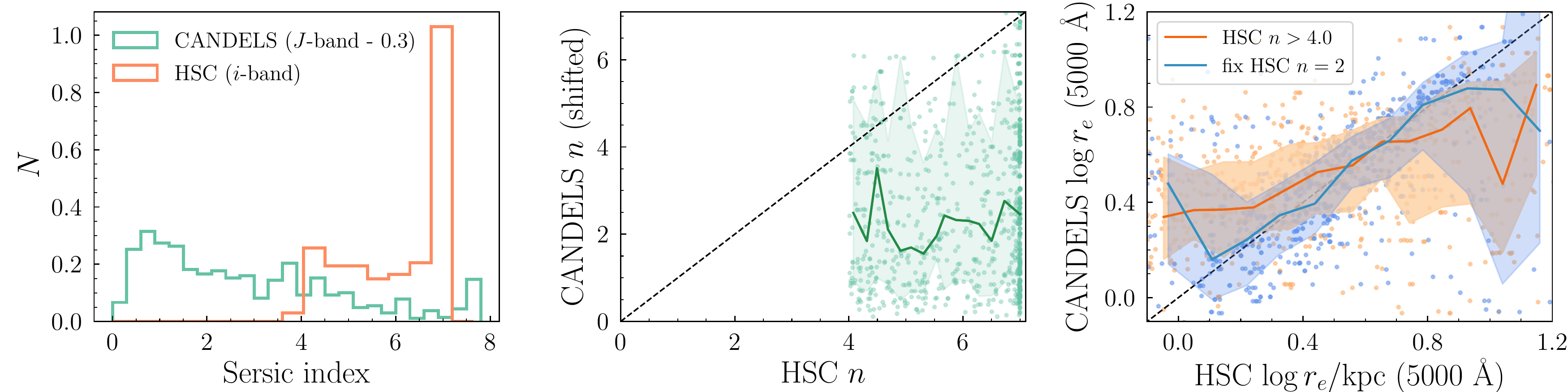}
\caption{Simulation results (III) using model quasars with real inactive CANDELS galaxies. For illustration, the sample is split into those with $n<4.0$ (top) and $n>4.0$ (bottom) based on the HSC values. Left column: distributions of the fitted (i.e., HSC) and intrinsic (i.e., CANDELS) \ss~indices for simulated quasars. Middle column: comparison of fitted and intrinsic \ss~indices. Green points show individual quasars while the curve and shaded region mark the 16-50-84th percentiles. Right column: comparison of fitted and intrinsic sizes for simulated quasars. The orange points show individual quasars while the orange curve and shaded region indicate the 16-50-84th percentiles. The sizes derived by fixing $n=2.0$ for simulated quasars with initial fit values of $n>4.0$ are shown in blue (see text).}
\label{fig:hitn}
\end{figure*}

We further assessed the reliability of our results and made appropriate adjustments to our method based on a comparison between HSC-based measurements and those derived from high resolution \hst~images and multi-band photometric data from UV to IR in the CANDELS-COSMOS field. As mentioned in Section \ref{subsec:simu}, our motivation was to evaluate uncertainties resulting from the use of simulations having idealized \ss~galaxy profiles, isolated quasars, and structural parameters fixed to the \i-band values for the other HSC bands. Here, we created simulated quasar images by inserting a point source (model HSC PSF) into HSC images of real CANDELS galaxies. We then compared the decomposition results to similar measurements of CANDELS galaxies prior to the addition of a model quasar. The structural and photometric properties of the CANDELS galaxy sample are derived from the data set and methods as described in Section \ref{subsec:control}. The galaxy properties provided by the CANDELS team should be more robust given the higher resolution of HST, wider wavelength coverage for SED fitting, and the fact that the measurements are free of quasar emission (although they may have their own measurement issues; \citealt{Haussler2007, Carollo2013}).

Specifically, we cross-matched the CANDELS data\footnote{https://3dhst.research.yale.edu/Data.php}, available in the COSMOS field, with the HSC DR2 Wide survey to build a sample of inactive galaxies. The HSC cmodel magnitudes in $grizy$ bands for these CANDELS galaxies are available as described in \citet{Aihara2019}, and we consider these as the true galaxy magnitudes in what follows.  We created cut-out HSC images and PSF models for 1141 matched sources with $0.2<z<1.0$ (328 of which have spectroscopic redshifts), $9.5 < {\rm log}\,M_*/\msun < 11.5$ and i\_cmodel\_mag~$< 23.0$. For each galaxy in CANDELS, we randomly selected a $M_*$- and redshift-matched SDSS quasar to identify an appropriate PSF model that is then added to the HSC image of the CANDELS galaxy in all five bands to create a series of simulated quasar+host images. The input flux density of the point source in the \i~band is matched to the actual value of the quasar and scaled appropriately to the effective wavelength of the other HSC filters using $f_\lambda = \beta * (\lambda/3000)^\alpha$, where $\alpha$ is the slope of the power-law fit to the quasar continuum given by \cite{Rakshit2019}, and $\beta$ is the normalization determined by the decomposed $i$-band quasar magnitude derived in Section \ref{subsec:decomp_method}. We then decomposed the composite image into the point source and galaxy components to measure the structural properties and pure galaxy fluxes of the underlying CANDELS galaxies as was done for the SDSS quasars.

We repeated the above procedure four times for each CANDELS galaxy and built a sample of $\sim5000$ simulated quasars. This procedure resulted in a simulated distribution of host-to-total flux ratio (\fr) that matched the observed quasar sample; the mean of the simulated distribution is only $\sim7\%$ smaller than the true value. Therefore, the simulation results presented here can be generalized to evaluate our decomposition and calibration results for SDSS quasars. 

In Figure \ref{fig:cmp_mag} we show the offsets between our decomposed galaxy magnitudes and the true galaxy magnitudes (i.e., decomposed magnitudes $-$ HSC cmodel magnitudes) as a function of the true galaxy magnitude. The data are binned into three ranges of the best-fit HSC \ss~index. For sources with HSC $n<4.0$, the decomposed host magnitudes are slightly offset ($\sim0.1-0.2$ mag) from the true galaxy magnitudes, consistent with the trend of underestimating the host-galaxy flux as reported in Section \ref{subsec:simu}. However, for those with $n>4.0$, our decomposed galaxy magnitudes deviate significantly from the true values. 

\begin{figure*}
\centering
\includegraphics[width=\linewidth]{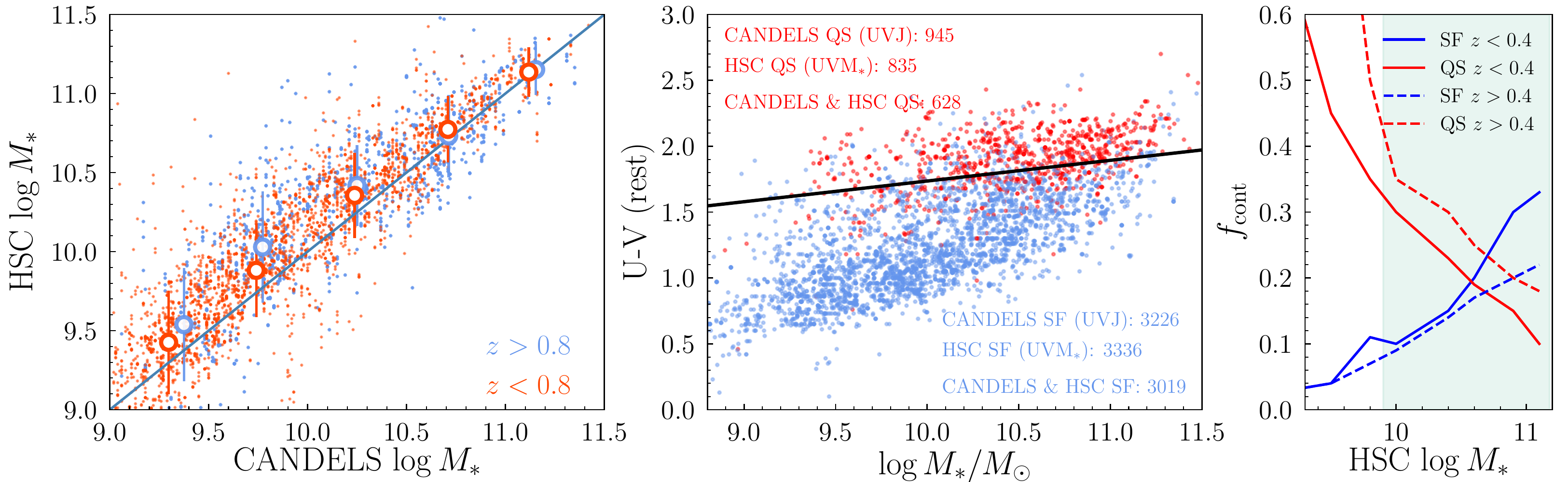}
\caption{Simulation results (IV): assessing the reliability of our SED fits by comparing the HSC- and CANDELS measurements: stellar mass (left panel) and rest-frame color vs. stellar mass (middle panel; see text in Section \ref{subsubsec:cmp_sed} for details). In the left panel, filled data points and empty circles represent individual sources and mean values (with standard dispersion), respectively. The right panel plots the contamination fraction of star-forming and quiescent quasar host galaxy classification as a function of \m, at redshift above and below 0.4. The $10-90$th percentile range of the stellar mass distribution for our SDSS quasar sample is indicated as the shaded region. }
\label{fig:candels}
\end{figure*}

\begin{figure}
\centering
\includegraphics[width=\linewidth]{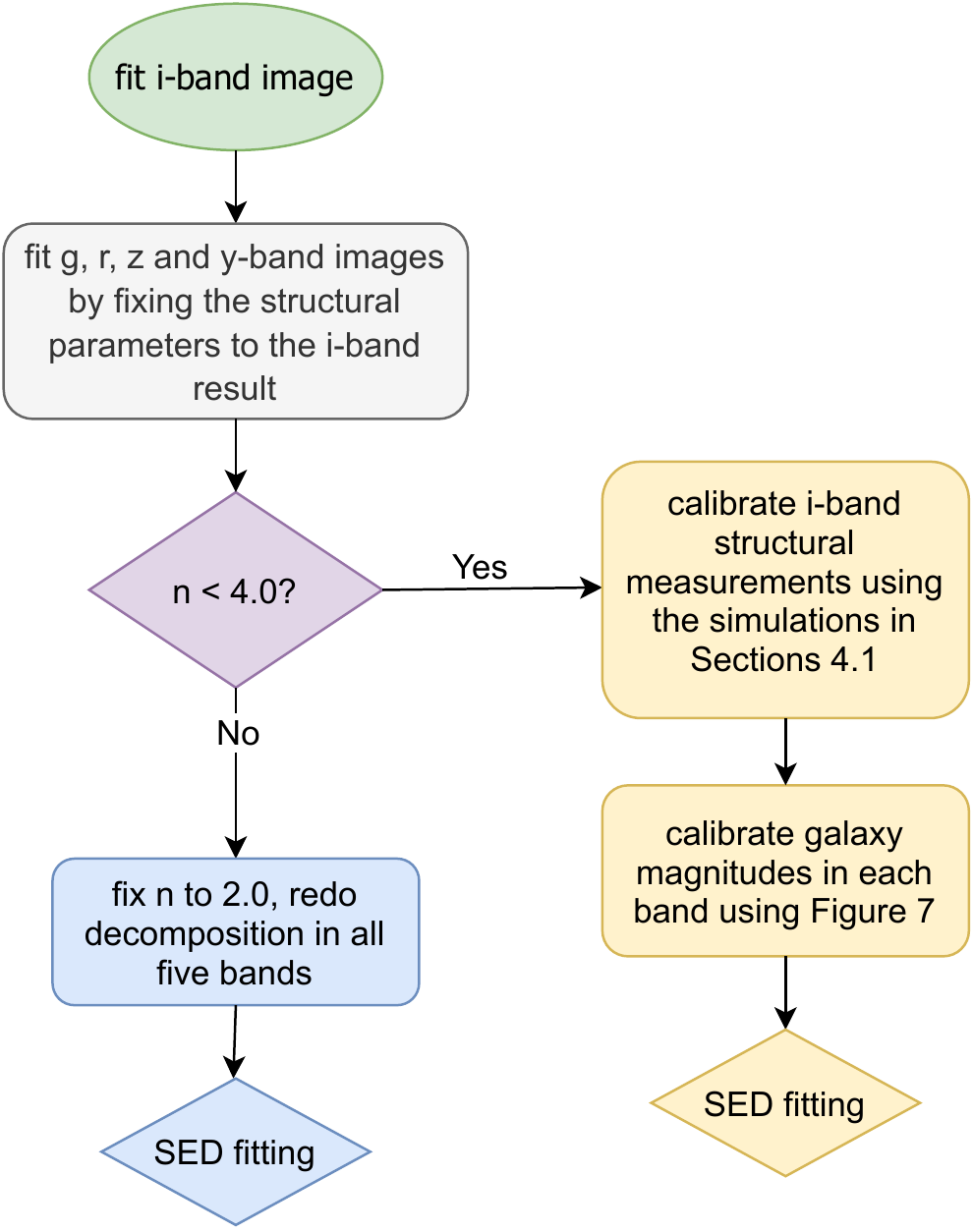}
\caption{Flow chart of the decomposition,  calibration and SED fitting procedure.}
\label{fig:flow}
\end{figure}

\subsubsection{Issues with Large \sid}
\label{ref:nlarge}
To further understand the origin of the large offset seen in Figure \ref{fig:cmp_mag} for galaxies with $n>4.0$, we compared our HSC-based \sis~and galaxy sizes with the corresponding values from HST in Figure \ref{fig:hitn}. 
The HSC $i$-band size and \sid~are calibrated using the image simulations presented in Section \ref{subsubsec:calibration}. The sizes derived from both HSC and HST images have been corrected to a common rest-frame wavelength of 5000~\ang~using the wavelength dependencies reported by \cite{vanderwel2014}. The HST-based \ss~indices are measured in the $J$-band, thus we systematically shift them by $-0.3$ \citep{Vulcani2014} to roughly match the \i-band result. As shown in the left and middle panels of Figure~\ref{fig:hitn}, the majority of simulated quasars have HSC \sis~less than 4.0 (which is also true for SDSS quasars; see Section \ref{subsec:sersic_index}) and trace the intrinsic values, thus recovering the distribution of true \sis. However, the large scatter demonstrates an inability to accurately recover the individual \sis~when using more realistic image simulations that include additional components (e.g., galaxy substructures, disturbed morphologies, blends with nearby objects, and other features). 
We do find that the sizes derived from HSC images (right column) are in excellent agreement with those derived from HST images, with the exception that the intrinsic sizes tend to be slightly higher ($\sim0.2$~dex) for HSC sizes smaller than $\sim1.5$~kpc. Fortunately, only $\sim4\%$ of our SDSS quasar sample have $\re<1.5$~kpc, thus this bias has a negligible impact on our results. 

For objects in which the fitted HSC \ss~indices are larger than 4.0, the HSC \ss~indices and sizes are poorly correlated with the intrinsic values. In fact, most of them are actually low-$n$ systems with a mean of 2.0 (bottom middle panel in Figure~\ref{fig:hitn}). The point source components in these simulated quasars tend to be under-subtracted, causing overestimates of the \sid~and host-galaxy fluxes. To calibrate the offset in galaxy magnitudes in an average sense, and to obtain more reasonable size measurements for the $n>4.0$ cases, we enact the following adjustment to our decomposition procedure. For sources with $n>4.0$ (HSC), we fix the \sid~to 2.0 (i.e., the mean of their true \ss~indices) and re-run the decomposition in all five bands. For those with $n<4.0$, we correct for the median deviation between our decomposed magnitudes and the true galaxy magnitudes in each band, which is derived as a function of our decomposed host magnitudes and the \sid. The resulting galaxy magnitudes and sizes are in better agreement with the true values, as can be seen from the bottom panels in Figure \ref{fig:cmp_mag} and the bottom right panel in Figure \ref{fig:hitn}. However, in the cases where we fix \sid~to 2.0, small offsets in galaxy magnitude ($\sim0.1-0.2$ mag) still remain for simulated quasars that have hosts brighter than 21~mag; this affects the stellar mass by only $\sim0.1$~dex for a small fraction ($10\%$) of our sample, leaving our overall result unaffected.

\subsubsection{SED Fitting}
\label{subsubsec:cmp_sed}
An SED fitting procedure is performed using the calibrated galaxy fluxes to derive photometric properties of the underlying CANDELS galaxies. Comparisons between HSC- and CANDELS-based $M_*$ and rest-frame $U-V$ vs. $M_*$ are shown in Figure \ref{fig:candels}. The HSC-based stellar-mass estimates are on average in good agreement with the CANDELS-based measurements with a small systematic offset at low stellar masses ($\sim0.1$~dex), at least up to $z\sim0.8$. The systematic offset could partly arise from systematic variations in the photometry in different surveys; the HSC cmodel magnitudes are typically $\sim0.1-0.2$~mag brighter than those given by the CANDELS catalog in the $grizy$ bands (see also K21). Throughout the following analysis, we directly used the stellar masses derived from CIGALE, and do not apply corrections for this offset to our stellar mass estimates while bearing in mind these difference when discussing our results.

The rest-frame $U-V$ colors are in good agreement with the CANDELS values, with a median difference of 0.03. We use the $U-V$ vs. $M_*$  diagram to separate quasar hosts into star-forming and quiescent galaxies as shown in Figure \ref{fig:candels}~(middle panel). The positions of individual points represent the HSC-based $UVM_*$ classification, and their colors are based on the true classification results (red for quiescent and blue for star-forming). The galaxy types classified by each criteria as well as the amount of consistent classifications between HSC and CANDELS are labeled.

To be clear, we adopt the $UVJ$ classification result based on CANDELS photometry as the true classification. In this regard, we are treating the $UVJ$ classification as a ground truth given the SED fitting result can be better constrained with the CANDELS dataset, although it has its own limitations on separating QGs from dusty SFGs \citep{Moresco2013}. Most of the mis-classified sources are close to the selection wedge, causing cross-contamination between the two populations. 

Figure \ref{fig:candels} (right panel) shows the contamination fraction, $f_{\rm cont}$, for the HSC-based identification of quiescent and star-forming host galaxies as a function of stellar mass. For the quiescent population, $f_{\rm cont}$ decreases with increasing stellar mass and is typically $\sim10-40\%$ for the mass range in which the majority of SDSS quasars lie. The contamination fraction for the star-forming population is typically $\sim10-30\%$ and increases with stellar mass. Overall, our simulation shows that we can successfully identify 89\% of star-forming hosts with an accuracy of 86\% based on HSC photometry for massive hosts ($\logm>10.0$). Since the fractions of quiescent and star-forming galaxies in the simulation are similar to what we observed for the SDSS quasar sample (see Section \ref{subsec:sf_prop}), we directly apply the estimated $f_{\rm cont}$ to SDSS quasars when discussing our result.

\begin{figure}
\centering
\includegraphics[width=\linewidth]{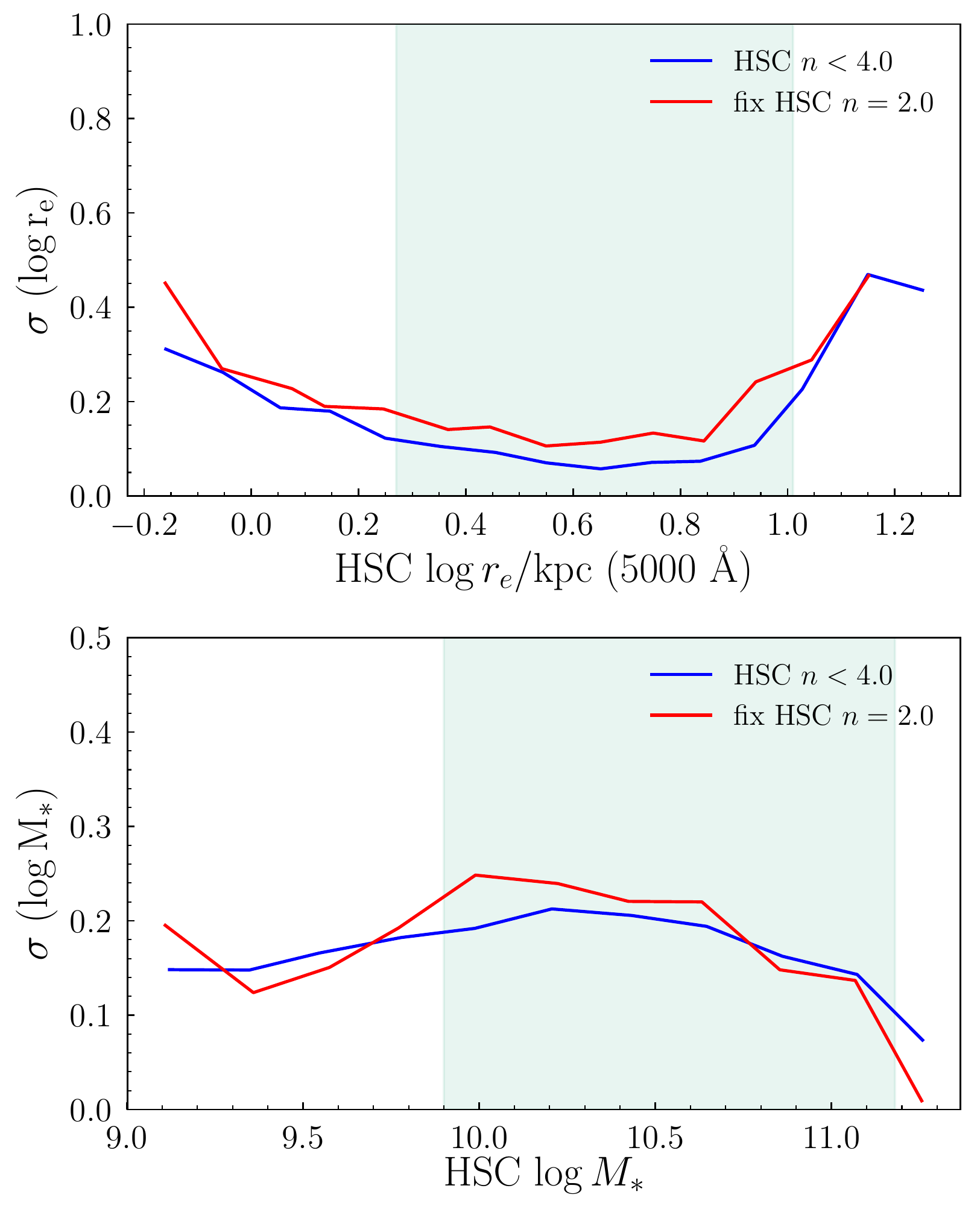}
\caption{Typical uncertainties in the measured galaxy sizes and stellar masses from quasar-host decomposition based on HSC imaging data, as estimated through the simulation described in Section \ref{subsec:evaluate}.}
\label{fig:error}
\end{figure}

Motivated by the simulations presented here, we perform the same calibration to SDSS quasars followed by an SED fitting to derive more reasonable structure and photometric properties for their host galaxies. The flow chart in Figure~\ref{fig:flow} summarizes our final decomposition, calibration and SED fitting procedures.
Since the measurement errors given by \lens~are unrealistically small, we adopt the typical scatter\footnote{To minimize the effect of outliers, we adopt the so-called ``robust statistical estimators'', i.e., the biweight location and scale to represent the median and the standard deviation of a distribution in the following analyses.} between HSC and CANDELS measurements as a more realistic estimation of errors for SDSS quasars. Figure~\ref{fig:error} summarizes the derived errors for galaxy sizes and stellar masses. The estimated errors on stellar mass are similar to those given by CIGALE. The error on $U-V$ color is typically 0.2~mag.

\begin{figure*}
\centering
\includegraphics[width=\linewidth]{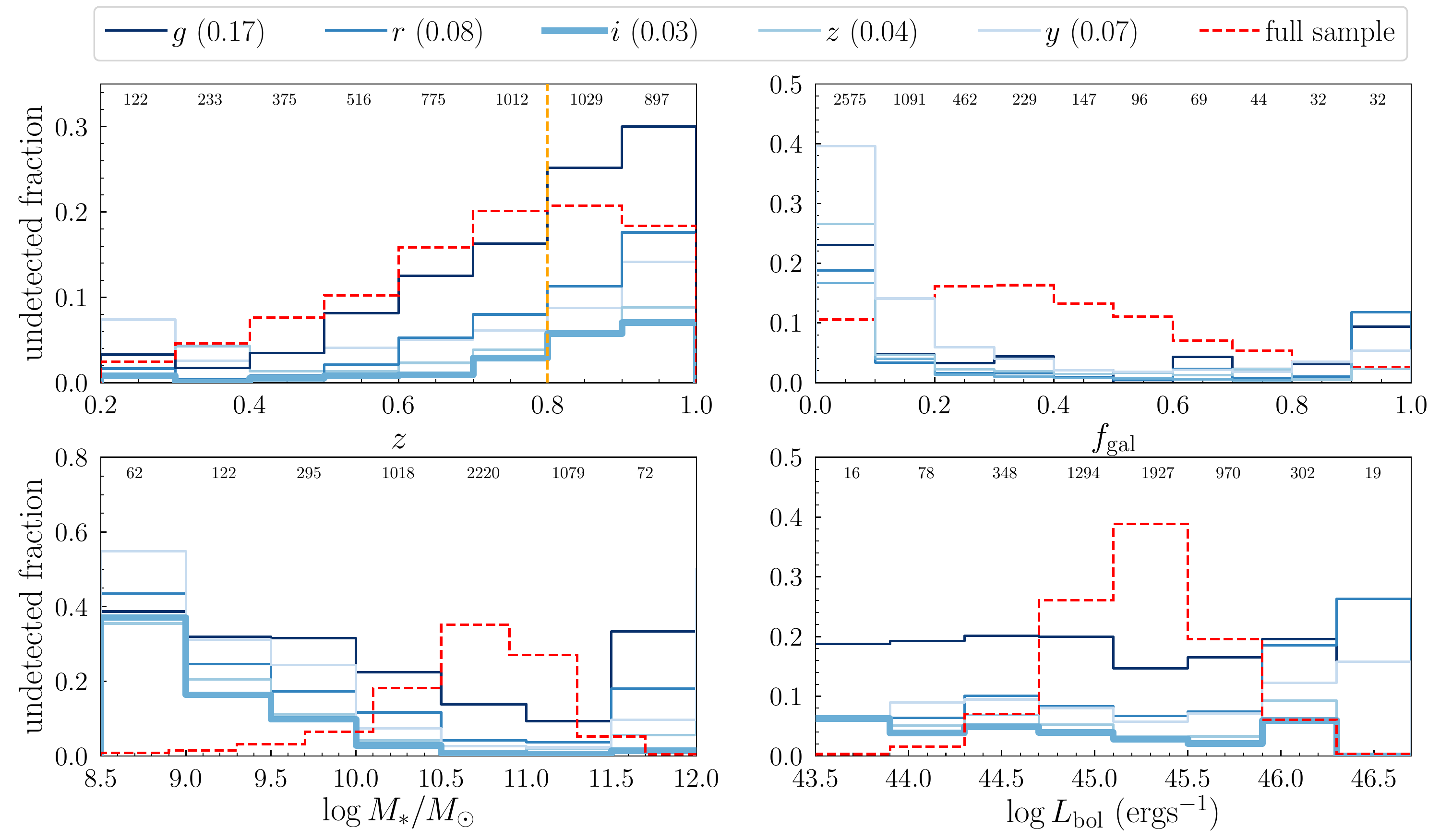}
\caption{Distributions of undetected fractions of quasar host galaxies as a function of redshift, host-to-total flux ratio, stellar mass and quasar bolometric luminosity in $grizy$ bands.  These parameters are the original fitted values before applying calibration corrections. 
The total number of sources in each parameter bin (\fr~for the \i-band only) are shown near the top of each panel. The numbers on the top of the image shown in parentheses represent the total fraction of undetected hosts in each band. The parameter distributions of the full quasar sample are shown in red. The vertical orange dashed line marks our redshift cut.}
\label{fig:detect_host}
\end{figure*}

\section{Final Sample Selection}
\label{sec:final_sample}

We set a stellar mass cut ($M_{*,\rm cut}$) to construct a relatively mass-complete sample that avoids problematic mass measurements due to large undetected fractions of quasar hosts at low stellar masses (see Section~\ref{sec:undetect}). The mass cut is chosen to correspond to the 90\% mass completeness limit at an \i-band magnitude of 23~mag using the method described in \cite{Pozzetti2010}. As a reference, the mass limits at $z=0.3$, 0.5 and 0.7 are $10^{9.3}\,M_\odot$, $10^{9.8}\,M_\odot$ and  $10^{10.3}\,M_\odot$, respectively. Quasars with $\logm>11.5$ are also excluded because of large undetected fractions and uncertain mass estimates (see Section~\ref{sec:undetect}). We also impose a requirement on the goodness-of-fit to select sources with reasonably accurate structural and photometric measurements. Our final sample selection criteria are summarized as follows:

\begin{itemize}[label={\tiny\raisebox{1ex}{\textbullet}}]
\setlength{\itemsep}{0pt}

\item $z<0.8$ (Section~\ref{sec:undetect})

\item host is detected in the \i-band with $\rm \Delta BIC_i > 10$ (Section~\ref{sec:undetect})

\item stellar mass is above the mass cut given above and below $10^{11.5}\,M_\odot$

\item reduced $\chi^2$ value of the SED fit is smaller than 10

\item the $i$-band $\Delta \rm SB$ value that represents the goodness-of-fit in the central quasar region (see Section \ref{subsec:decomp_method}) is smaller than 0.03

\item the $\omega$ value, representing the reliability of the fitting results and applied corrections (see Section \ref{subsubsec:calibration}), is smaller than 2.0 for sources that don't have a fixed \sid

\item at least two model galaxies can be found to derive corrections (see Section \ref{subsubsec:calibration}) for sources without a fixed \sid.

\end{itemize}

The last four criteria are set empirically to exclude sources with poor or unreliable fits. Strengthening or relaxing these criteria slightly does not significantly affect our results. The number of sources rejected by each criteria are summarized in Table \ref{table:exclude}. For the 2976 quasars at $z<0.8$, 436 of them are excluded because of the stellar mass cut. An additional 21 sources are excluded since their hosts are undetected in the \i-band (Section~\ref{sec:undetect}), and another 95 sources are further rejected because of poor fits. The final sample contains 2424 sources which are exclusively used for the analysis in Section~\ref{sec:results} and following discussion (Section~\ref{sec:discussion}). With our redshift and stellar mass cut, 90\% of the remaining sources have $\fr>0.2$, thus we are able to reliably perform quasar-host decomposition for the majority of sources in our final sample. For the rejected sources that are either undetected or have poor or unlikely fits, we have little knowledge about their intrinsic structural and stellar population properties, although most of them are likely to be faint or compact systems. However, those objects only account for 3\% of our sample, thus excluding them will not significantly influence the results. 

\subsection{Undetected Host Galaxies} 
\label{sec:undetect}
In Figure \ref{fig:detect_host} we show the fraction of quasars with non-detections for their host galaxies as functions of $z$, \fr, $M_*$ and \lbol~in each band. The parameters plotted in Figure \ref{fig:detect_host} are the original fitted values  before applying calibration corrections. 
We define the host to be detected if $\Delta \rm BIC > 10$ (Section \ref{subsec:decomp_method}). The non-detection fractions are 17\%, 8\%, 3\%, 4\% and 7\% for the \g, \r, \i, \z\ and \y~band, respectively. The \i-band shows the highest detection fraction (97\%), due to the better seeing condition and the relatively high host-to-quasar contrast. The lowest detection fraction occurs in the \g-band, which can be up to 30\% at $z>0.8$. Most of the undetected sources are at high redshifts due to surface brightness dimming, and have  lower values of \fr~and \re, due to the issue of resolving faint or/and compact galaxies (see Section \ref{subsec:evaluate}). Massive galaxies ($\logm>11.5$) at high redshifts tend to have lower detection rates in bluer bands because (1) they are mainly quiescent systems, thus their rest-frame emissions which are below the 4000~\ang~break are weak and (2) the measured flux of undetected host could contain part of the quasar emission thus giving us a wrong stellar mass. About 5\%--10\% of quasars in each band with fitted \fr~larger than 90\% also have undetected hosts, likely due to the \lens~fit mistakenly treating the unresolved quasar as a bright compact galaxy. For quasars with undetected hosts, we are unable to reliably recover their intrinsic structural parameters, thus we require the host to be detected in the \i-band. The large non-detection fractions in the \g-band and \r-band at $z>0.8$ and $\logm>11.5$ also prevent us from deriving reliable SED fitting results. Therefore, we restrict the sample to quasars at $z<0.8$ and $\logm<11.5$.

\begin{figure*}
\centering
\includegraphics[width=\linewidth]{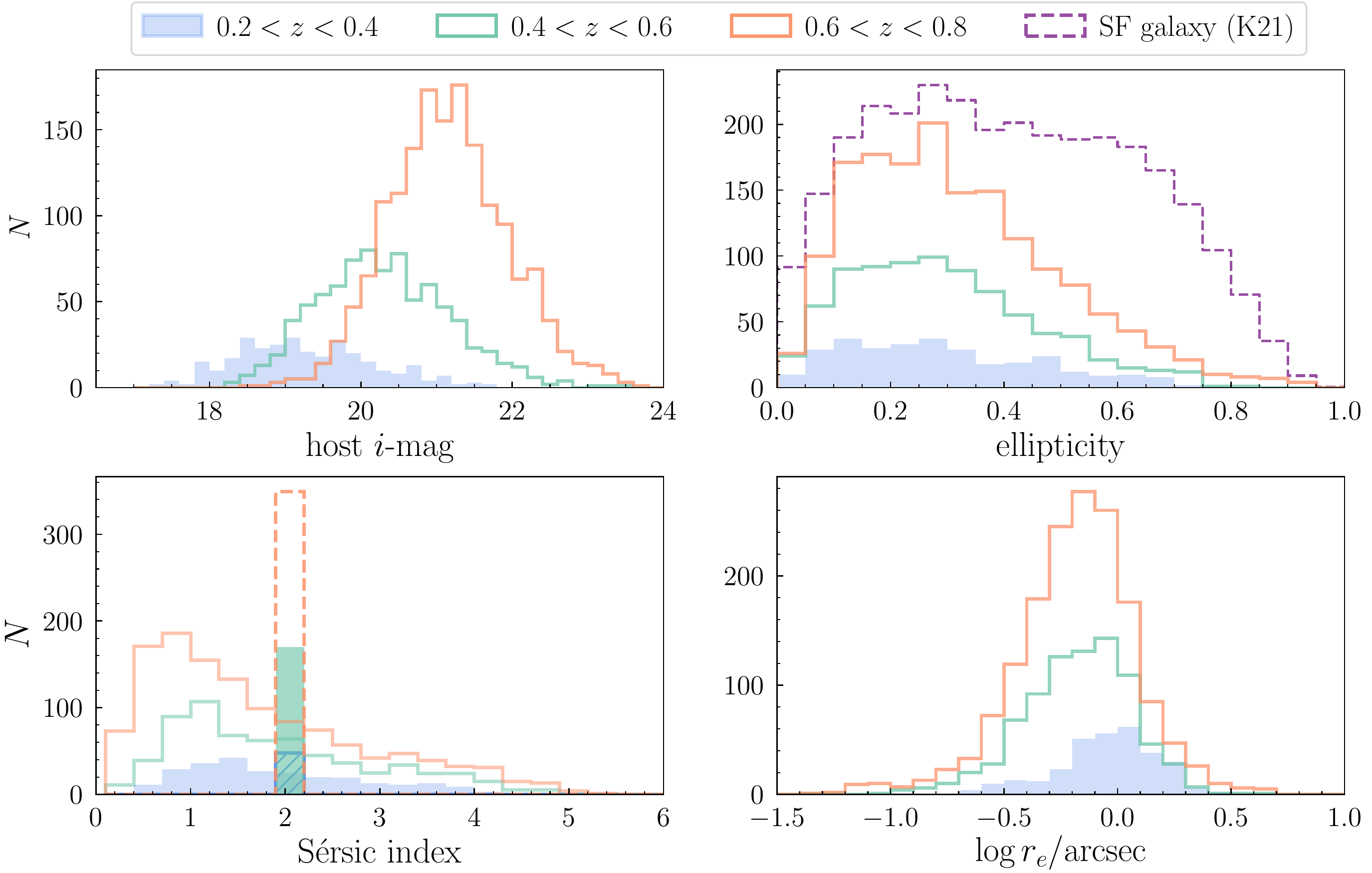}
\caption{Distributions of the calibrated host-galaxy magnitude (top left), ellipticity (top right), \sid~(bottom left) and size (bottom right) for the final sample as measured in the \i~band. The narrow bin at $n=2$ in the bottom left panel includes sources with their \ss~indices fixed.
}
\label{fig:params}
\end{figure*}

\section{Results}
\label{sec:results}
Figure \ref{fig:params} shows the calibrated distributions of the decomposed host-galaxy magnitude, ellipticity, \sid~and size measured in the \i-band for our final sample in three redshift ranges. The majority of our sample have hosts brighter than 22~mag, $\re<1.5\arcsec$ and $n<3.0$. Thus, we can constrain their stellar masses and sizes well, as demonstrated using the simulations we presented in Section \ref{sec:simu}.

\subsection{Flux of Quasar Host Galaxies}
\label{subsec:host_ratio}

\begin{figure}
\includegraphics[width=\linewidth]{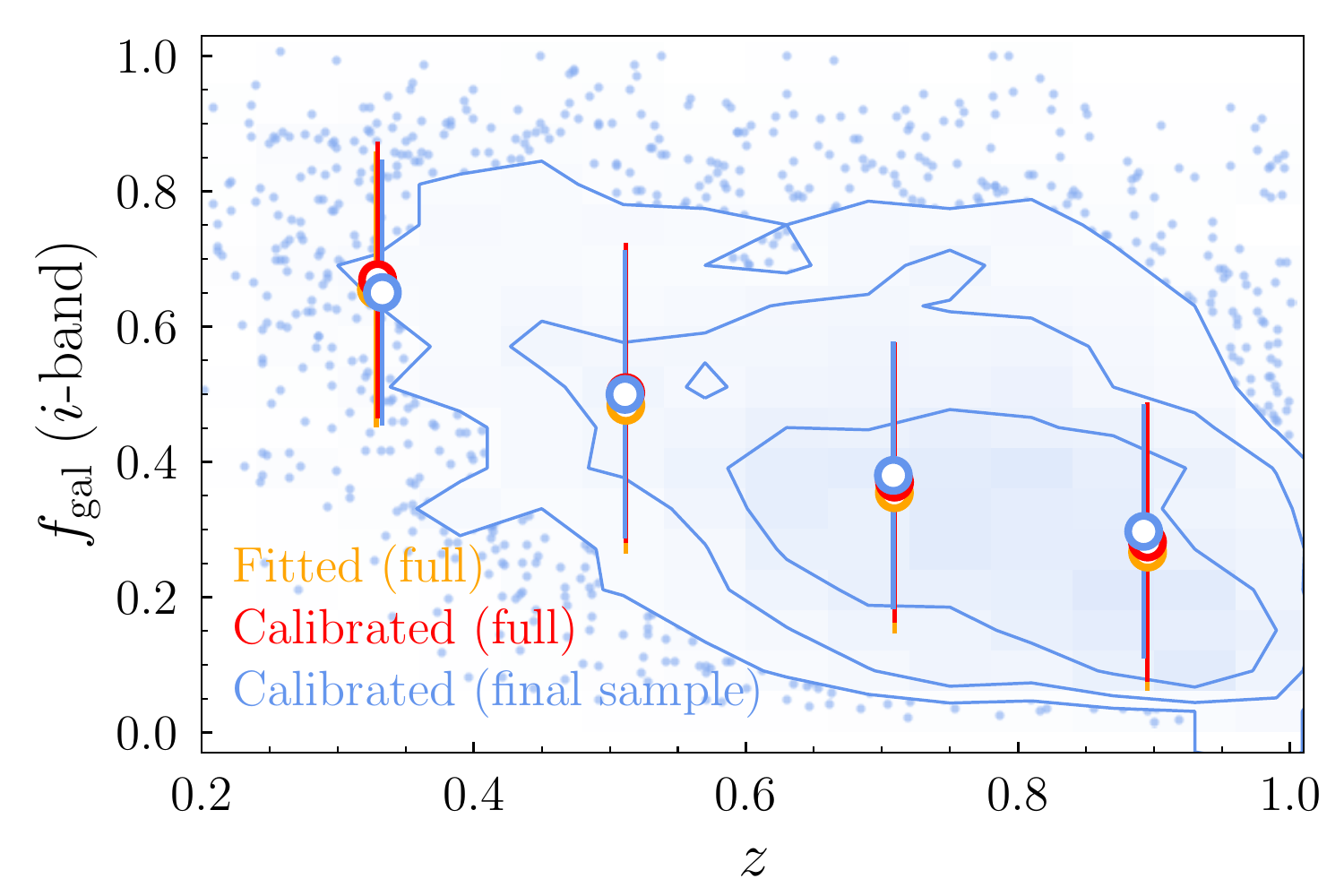}
\caption{The \i-band host-to-total flux ratio as a function of redshift. The median values in four redshift bins are shown as empty circles with the error bars representing the dispersion.}
\label{fig:host_frac}
\end{figure}

Figure \ref{fig:host_frac} shows the fitted and calibrated \fr~in the \i-band as a function of redshift. Sources at $z>0.8$ are also shown to better illustrate the trend. 
The \fr~is high overall (typically larger than 20\%), thus the host galaxy has a significant contribution to the total emission in the optical band over the full redshift range considered here.
While the selection procedure tends to exclude sources with faint hosts, it only causes a negligible change on the median values of \fr. 
The median \fr~decreases from about $\sim70\%$ at $z\sim0.2$ to $\sim30\%$ at $z\sim1$, and the detection fraction also drops at high redshift (see Figure \ref{fig:detect_host}), consistent with the trends reported in \cite{Ishino2020}. The declining trend is due to the combination of the differential surface brightness dimming between the host ($\propto (1+z)^{-4}$) and the quasar ($\propto (1+z)^{-2}$), the fact that a given filter probes bluer rest-frame wavelengths at higher redshifts where quasar emission usually outshines that of its host galaxy, and that the quasars are of higher luminosity at higher redshift in a flux-limited survey such as SDSS. 

\subsection{Ellipticity of Quasar Host Galaxies}
\label{subsec:ellp}
Figure \ref{fig:params} (top-right panel) shows the distribution of ellipticity ($1-b/a$) for the host galaxies of SDSS quasars, where $a$ and $b$ are the semi-major and semi-minor axes from our 2D modelling. Overlaid are the \m-matched inactive SFGs from K21. The ellipticity distributions of SDSS quasars are clearly shifted to lower values. Assuming that the underlying 3D shapes of quasar hosts are the same as those of inactive galaxies of similar types (see Section \ref{subsec:sf_prop}), our result suggests that SDSS quasars are preferentially biased toward face-on systems, while inactive galaxies have an ellipticity distribution that is more evenly distributed \citep{Lagos2011, Malizia2020}. 
If the orientation of AGN accretion disk is misaligned with the galaxy disk, as supported by previous observations by using the radio jet as a tracer of the direction of  accretion disk \citep{Kinney2000, Schmitt2002}, the lack of edge-on hosts among SDSS quasars would indicate that the galactic-scale dust can also play a role in obscuring the broad emission lines \citep{Maiolino1995, Goulding2009, Goulding2012}. This is inconsistent with the AGN unification model, where the AGN types are exclusively determined by the inclination angle relative to the obscuring torus/accretion disk.
Alternatively, the direction of galaxy disk may be co-aligned with the AGN accretion disk and nuclear torus to some level \citep{Battye2009}, leading to a deficit of edge-on galaxies hosting type-1 quasars.

\subsection{$S\acute{e}rsic$ Indices of Quasar Host Galaxies}
\label{subsec:sersic_index}
There has been much effort in the literature to determine whether AGNs prefer to reside in a galaxy of a particular type (e.g., disk- or bulge-dominated, star-forming or quiescent). To address this issue, we plot the distribution of the \sid~in Figure \ref{fig:params}~(bottom-left panel) with values calibrated as described in Section~\ref{sec:simu}. As a reminder, the \sid\ is a parameteric measure of the radial light distribution that has been shown to discriminate between disk- and bulge-dominated galaxies. 
We first point out that the sharp peak at $n=2.0$ is artificial since most galaxies in this bin had their \sid~held fixed (Section~\ref{ref:nlarge}).
Putting those objects aside, the \sid~distribution spans a broad range, with a peak value of 1.0 and a median value of 2.0. The overall low \ss~indices indicate that our quasars are hosted by galaxies with disk-like light profiles, consistent with previous studies on the morphologies of SDSS quasars \citep[e.g.,][]{Matsuoka2014, Yue2018} and X-ray-selected AGNs \citep[e.g.,][]{Schawinski2011, Kocevski2012, Silverman2019}. 

We highlight the fact that the \sid~of a quasar host is challenging to measure accurately. This is seen in the middle panels of Figure~\ref{fig:hitn} based on CANDELS galaxies and in Appendix~\ref{appendix:b} using simulations. However, the recovered \sis~significantly improve for those with higher values of $f_{gal}$ ($>0.2$) and brighter quasar magnitudes ($<21$ mag). Even so, our subsequent results, presented in this study, are focused on the measure of the effective radii of quasar host galaxies that are measured with higher precision.

\subsection{Star-forming Activity of Quasar Host Galaxies}
\label{subsec:sf_prop}

\begin{figure}
\centering
\includegraphics[width=\linewidth]{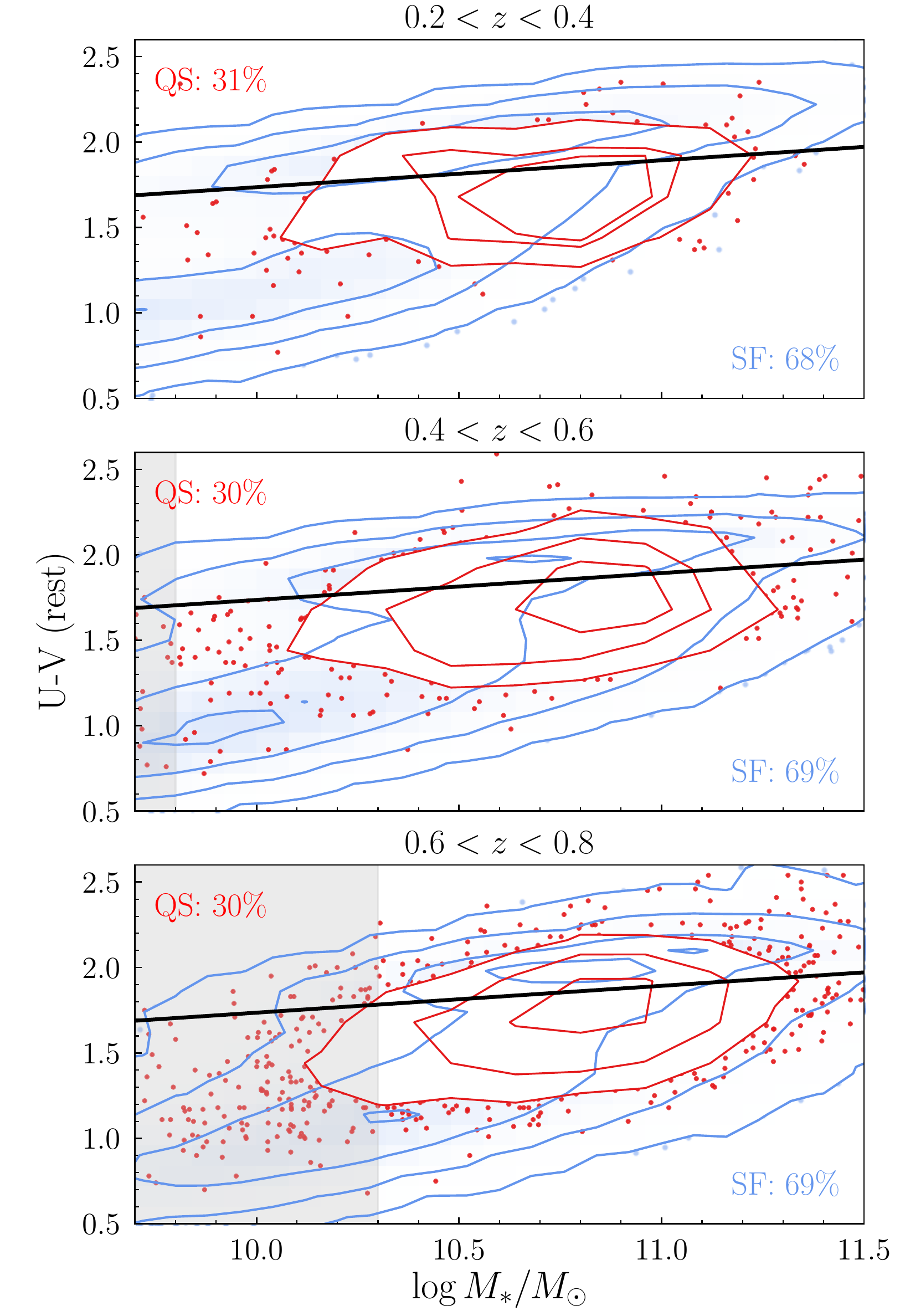}
\caption{Classification of quasar host galaxies as either star-forming or quiescent based on the rest-frame $U-V$ vs. $M_*$ diagram. Host galaxies of SDSS quasars are plotted as red contours and data points. Inactive galaxies from \cite{K21} are shown in blue. The vertical shaded region marks our \m~cut. The slanted black line indicates the division between galaxy types (SF, QS) with the fractions as shown.}
\label{fig:uvj}
\end{figure}

To evaluate the star-forming activity of our quasar hosts, we separate them into quiescent and star-forming galaxies using the rest-frame color $U-V$ versus $M_*$ diagram as shown in Figure \ref{fig:uvj}.  
While quasar hosts are widely distributed on the $U-V$ vs. stellar mass plane, the contours indicate that they are preferentially located in a region that is bluer relative to the red sequence, and more massive and redder than the high density region of blue galaxies. About 70\% of quasar hosts are classified as star-forming over the full redshift range considered. 
Therefore, we conclude that quasars are preferentially hosted by massive, star-forming galaxies. This result is in line with findings at similar or higher redshifts that luminous AGNs tend to be hosted by galaxies that are forming stars at substantial rates, either on the star-forming main sequence or undergoing a recent starburst \citep[e.g.,][]{Silverman2009, Mullaney2012, Rosario2013, Matsuoka2014, Schulze2019, Xie2021}. These results do not appear to support a scenario in which quasar feedback is in the process of suppressing or quenching star formation.

\begin{figure*}
\centering
\includegraphics[width=\linewidth]{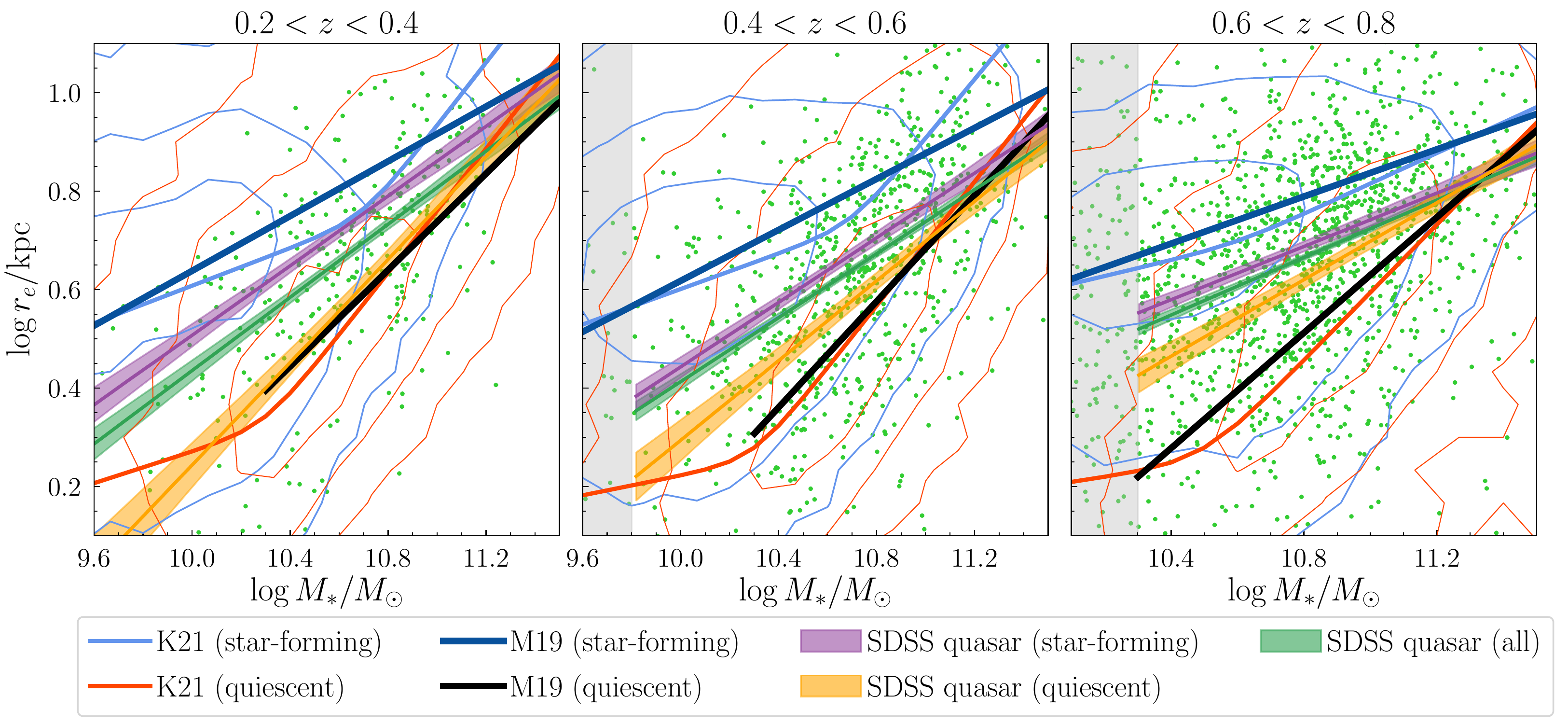}
\caption{Size -- stellar mass relation of quasar host galaxies. Small green points are the measurements for individual quasar hosts. Single-power-law fits to the \re--\m~relations are shown in green, purple and orange for all, star-forming, and quiescent host galaxies, respectively. Best-fit parameters are given in Table \ref{table:fit}. For comparison, we show the \re--\m~distributions for inactive SFGs (blue) and QGs (red) from \cite{K21} as contours with the outer one at the $3\sigma$ level. Double-power-law fits to their \re--\m~distributions are shown as blue (SFGs) and red (QGs) 
curves, respectively. In addition, we display the single-power-law fits to the \re--\m~relationship for inactive SFGs and QGs in \cite{Mowla2019} as dark blue and black lines, respectively. 
}
\label{fig:mass_size}
\end{figure*}

\begin{figure}
\centering
\includegraphics[width=\linewidth]{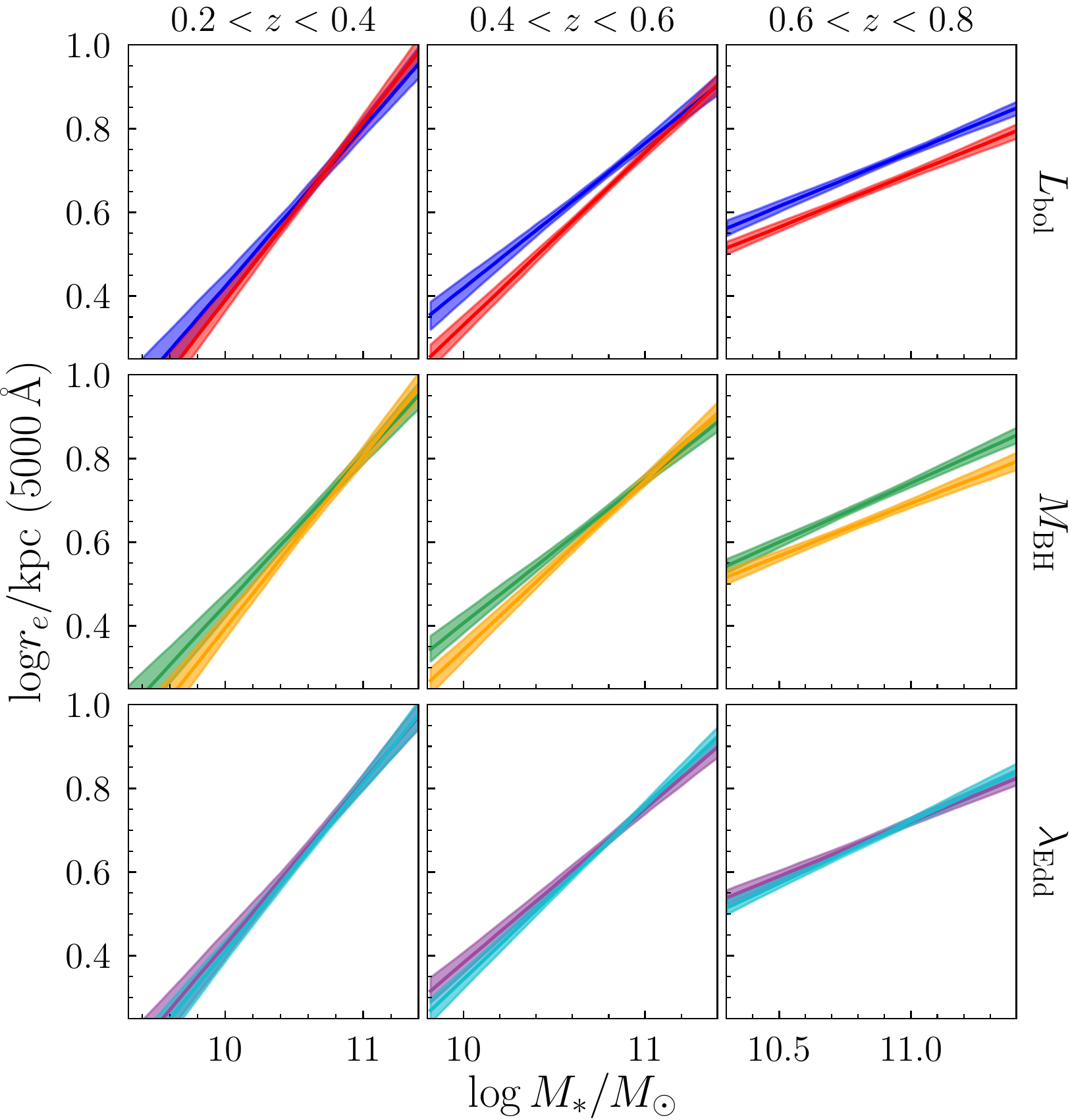}
\caption{Comparing the \re--\m~relation 
between quasars with high and low \lbol~(top), \mbh~(middle) and \edd~(bottom). Quasars with higher \lbol, \mbh~and \edd~are shown in blue, green and purple, respectively. No significant differences are found in the \re--\m~relation for particular quasar populations.
}
\label{fig:size_bin}
\end{figure}

\subsection{Sizes of Quasar Host Galaxies}
\label{subsec:size}

In Figure \ref{fig:mass_size}, we plot the sizes of quasar hosts as a function of \m~in three redshift bins. The individual measurements are shown as green points. A correlation of \re~with \m~can be clearly seen, albeit with large observational scatter. As detailed below, a positive linear relationship is found in each bin. 

To aid in our interpretation of these results, we also show the best-fit \re--\m~relations for inactive galaxies from M19 and K21. As a reminder, all sizes have been corrected to a common rest-frame wavelength of 5000~\ang. In particular, the \re--\m~relation given in M19 is represented as a single-power-law model, as shown by the dark blue (for SFGs) and black (for QGs) lines. K21 presented a \re--\m~relation for HSC galaxies that is described by a double-power-law for both SF and QGs. Interestingly, although quasars are widely distributed on the \re--\m~plane, a large fraction of them are located below the \re--\m~relations for inactive SFGs, especially considering that they are mainly hosted by SFGs.

With the large quasar sample from HSC and SDSS, we are able to fit the observed distribution of $\re$ and $\m$ with a parameterized model. We adopt a single-power-law model following \cite{vanderwel2014}.
We assume that the intrinsic size distribution at a given stellar mass follows a log-normal distribution $\mathcal{N}({\rm log\,}r, \sigma_{{\rm log}\,r})$, where ${\rm log}\,r$ is the mean and $\sigma_{{\rm log}\,r}$ is the intrinsic scatter. Additionally, $r$ is taken to be a function of stellar mass:
\begin{equation}
    r(\m)/{\rm kpc} = A \times m_*^\alpha,
\label{eq:pwl}
\end{equation}
where $m_* \equiv M_*/7\times10^{10}\,M_\odot$.
Assuming that the observed \re~has a Gaussian $1\sigma$ uncertainty of $\sigma({\rm log}\,r_e)$, the probability of observing a quasar with size \re~and mass $M_*$ is the inner product of two Gaussians:
\begin{equation}
    P = \langle \mathcal{N}({\rm log\,}r, \sigma_{{\rm log}\,r}),\ \mathcal{N}({\rm log\,}r_e, \sigma({\rm log}\,r_e)) \rangle.
\end{equation}
The uncertainties in \re~and \m~are shown in Figure~\ref{fig:error}, and we convert the error of \m~into an additional uncertainty on \re~assuming $\alpha=0.7$ for quiescent hosts and $\alpha=0.2$ for star-forming hosts. For quasars with a stellar mass error below 0.1 dex, we set it to 0.1 dex.
Similar to \cite{vanderwel2014}, we assign a weight $W$ to each galaxy that is inversely proportional to the number density using the stellar mass function from \cite{Muzzin2013}. This prevents the fit from being dominated by the large number of low-mass galaxies. 
We then derive the total likelihood for observing the SDSS quasar sample as a function of intercept $A$, slope $\alpha$ and intrinsic scatter $\sigma_{{\rm log}\,r}$:
\begin{equation}
    \mathcal{L} = \sum {\rm ln}\,(W*P).
\end{equation}

We also perform analytic fits to quasars hosted by SFGs and QGs separately while taking into consideration an assessment of the level of misclassification between the two populations. The fractional level of contamination $f_{\rm cont, Q}$ and $f_{\rm cont, SF}$ for each population is estimated based on our simulations as described in Section~\ref{subsec:evaluate} and shown in Figure~\ref{fig:candels}. In addition, we set $f_{\rm cont, SF}$ to zero if the observed $U-V$ color for a quasar is bluer than 1.0. We then compute the total likelihood for observing each type of host galaxy which can have a different intercept, slope and intrinsic scatter:
\begin{equation}
    \mathcal{L_{\rm Q}} = \sum {\rm ln}\,\lbrace W_{\rm Q} \cdot [(1-f_{\rm cont, Q}) \cdot P_{\rm Q} + f_{\rm cont, Q} \cdot P_{\rm SF}] \rbrace
\end{equation}
for quiescent hosts, and 
\begin{equation}
    \mathcal{L_{\rm SF}} = \sum {\rm ln}\,\lbrace W_{\rm SF} \cdot [(1-f_{\rm cont, SF}) \cdot P_{\rm SF} + f_{\rm cont, SF} \cdot P_{\rm Q}] \rbrace
\end{equation}
for star-forming hosts. We obtain the best-fitting parameters by maximizing the total likelihood, $\mathcal{L} = \mathcal{L_{\rm Q} + \mathcal{L_{\rm SF}}}$. We exclude objects that lie outside of the $2.5\sigma$ contour on the \re--\m~plane in the fitting to minimize sensitivity to outliers with artificially small errors in \re~and \m. 
After this cut, $\sim95\%$ of the objects in each redshift bin are used in the fitting. The best-fit \re--\m~relation as well as its $1\sigma$ confidence region are derived through MCMC sampling using emcee. 

In Figure \ref{fig:mass_size}, we display the fitting results for all quasars, star-forming quasars and quiescent quasars by green, purple and orange lines, respectively. The best-fit parameters are summarized in Table \ref{table:fit}. Quasar host galaxies show a clear \re--\m~relation, and their sizes on average tend to lie between those of inactive SFGs and QGs at equivalent masses from both K21 and M19. Note that the corrections we applied to the size measurements are small (0.03\arcsec~in the median), so the result is not driven by that. Systematically shifting our stellar mass measurements to correct for the average offset between our stellar mass and those of M19 (see Section \ref{subsec:evaluate}) does not affect this conclusion.

While quasar hosts can be either star-forming or quiescent, our result is not driven simply by a mixture of star-forming and quiescent hosts. The sizes of quasar hosts that are star-forming are still smaller than inactive SFGs at similar stellar masses. 
The $U-V$ color distributions for star-forming quasars and inactive SFGs are similar when controlling for stellar mass and redshift, thus our result is not driven by a surfeit of high central concentration green valley objects \citep[e.g.,][]{Mendez2011} among quasar hosts. 
This result is also robust to the position of the dividing line between SFG and QG in Figure \ref{fig:uvj}. 
Therefore, we find that quasar hosts are preferentially compact SFGs, in agreement with recent results at higher redshifts \citep{Kocevski2017, Silverman2019} and suggesting the existence of a physical mechanism whereby quasars are preferentially found in more centrally concentrated galaxies  \citep[e.g.,][]{Barro2014, Chang2017, Kocevski2017}.

The sizes for quasars in QGs are on average smaller than those hosted by SFGs, as expected. The \re--\m~relation for quiescent quasars follows that of inactive QGs at $z<0.4$. Beyond redshift of 0.4, the sizes for quiescent quasars are larger than those of inactive QGs at low stellar masses. However, we cannot rule out that this excess in size may be due to the fact that our model is not able to fully eliminate contamination from quasars with star-forming hosts.

To further investigate the dependence of galaxy size on quasar properties, we split our sample into different subsamples based on the median quasar bolometric luminosity ($\loglbol\sim45.2$), black hole mass ($\logmbh \sim 8.2$) and Eddington ratio ($\logedd \sim -1.2$) in a given redshift and stellar mass bin. The \re--\m~relations for different subsamples are plotted in Figure \ref{fig:size_bin}. We do not find significant difference in size (at most $\sim0.05$ dex in some bins) between quasars with high and low \lbol, \mbh~and \edd.

\begin{figure}
\centering
\includegraphics[width=\linewidth]{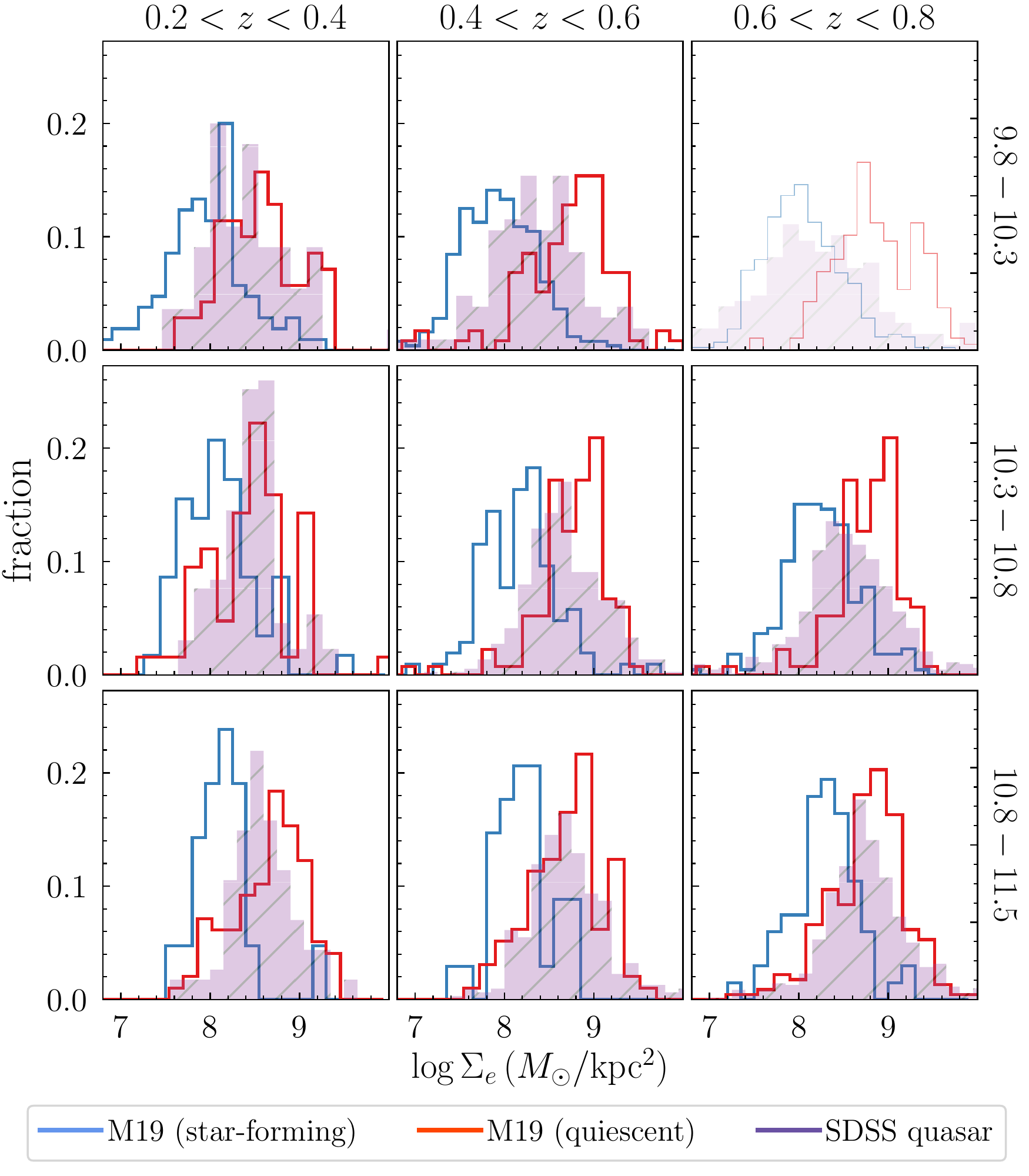}
\caption{Distributions of surface mass density of quasar host galaxies compared to that of inactive SFGs and QGs in \cite{Mowla2019}, shown in different redshift  (left to right) and stellar mass bins (top to bottom). The mass range below our stellar mass cut is drawn in light colors (top-right panel).}
\label{fig:sigma}
\end{figure}

\begin{figure}
\centering
\includegraphics[width=\linewidth]{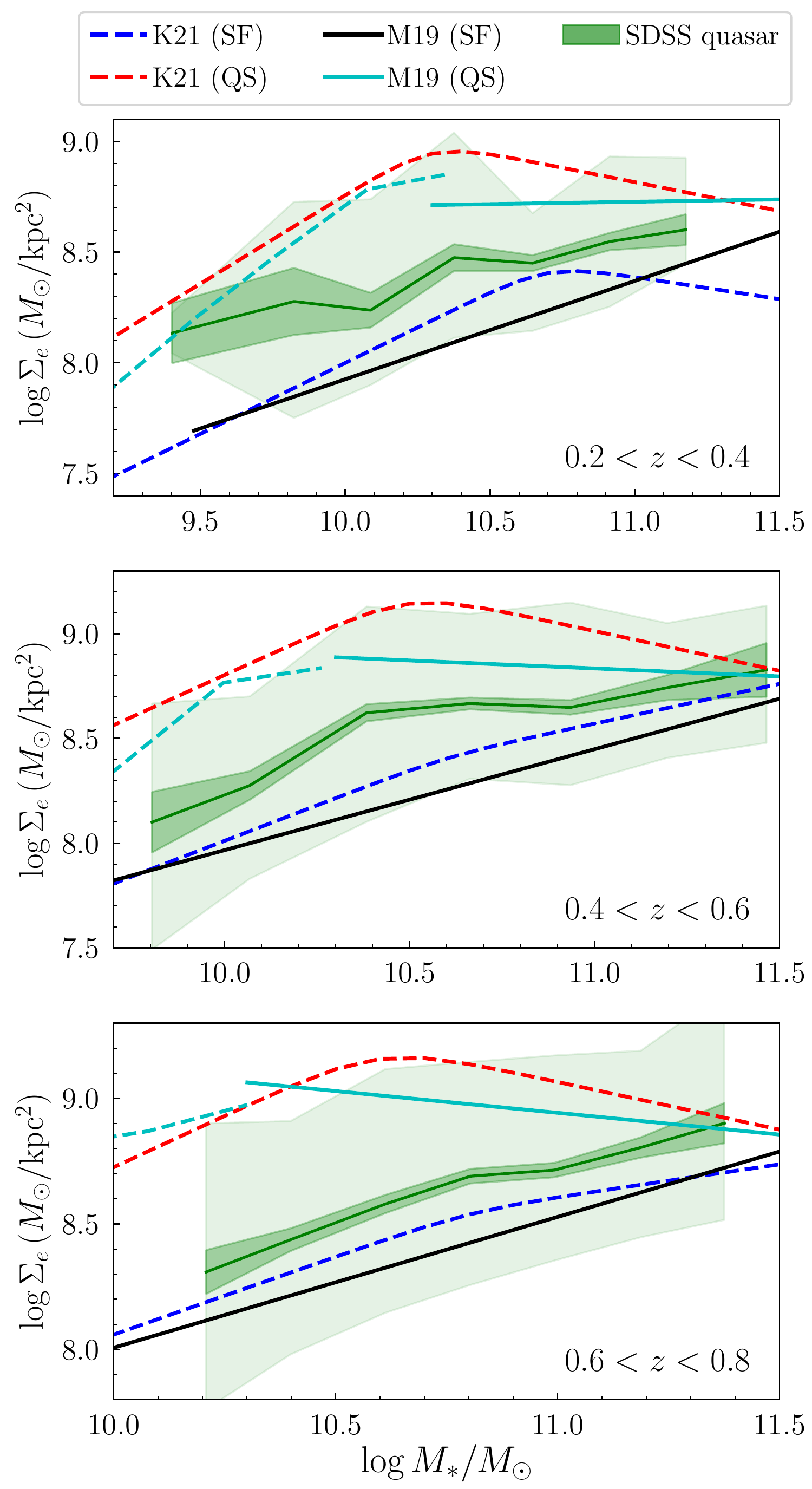}
\caption{Average surface mass density vs. stellar mass for SDSS quasars (green line, with the error on mean shown as dark green  region, and the 16--84th percentile range shown as light green region) compared to those of inactive galaxies in \cite{Mowla2019} and \cite{K21}. The \sigmae~for SDSS quasars increases with stellar mass, and is elevated relative to the general inactive star-forming populations. Combined with the result in Figure \ref{fig:uvj}, we conclude that quasars prefer to lie in compact SFGs.}
\label{fig:sigma_m}
\end{figure}

\subsection{Surface Mass Density of Quasar Host Galaxies}
\label{subsec:density}
Observationally, the SFH and quenching of a galaxy appear to be strongly connected to its compactness \citep{Kauffmann2003, Franx2008, Fang2013, Whitaker2017, Barro2017, Luo2020}, which is usually measured by the effective surface mass density ($\sigmae = 0.5\,M_*/\pi \re^2$) or the surface mass density within the central 1 kpc ($\Sigma_1$). In Section \ref{subsec:size}, we have demonstrated that quasar host galaxies exhibit a clear \re--\m~relation, which lies between those of the inactive galaxy populations. However, both size and stellar mass increase for massive galaxies, thus how the compactness of quasar hosts changes with \m~and compares to inactive galaxies are yet to be explored.

In Figure \ref{fig:sigma}, we plot the distributions of \sigmae~for both quasar hosts and inactive galaxies in M19. The \sigmae~distribution for quasar hosts does indeed lie between those of star-forming and quiescent galaxies, while the three populations have similar observed scatter. The \sigmae--\m~relations in three redshift intervals for quasar host galaxies are plotted in Figure \ref{fig:sigma_m}. Also shown are the \sigmae--\m~relations for inactive galaxies in M19 and K21 as derived from their best-fit \re--\m~relations. Note that due to the obvious flattening of the \re--\m~relation for low-mass quiescent galaxies in M19, they only carried out single-power-law fits for galaxies with $\logm>10.2$ (cyan solid curves). For those with $\logm<10.2$, we show their mean \re--\m~relation as cyan dashed curves. 

Inactive star-forming and quiescent galaxies exhibit clearly different patterns on the \sigmae--\m~plane. The surface mass density for quiescent galaxies peaks at the pivot mass ($\logm\sim10.6$), and decreases on both sides \citep{Barro2017}. The distinct patterns for quiescent galaxies with different stellar masses are likely reflecting different quenching and size growth mechanisms \citep{vanderwel2014}.
Star-forming galaxies, however, have lower \sigmae, and show an increasing trend with stellar mass; although the surface mass density slightly decreases at $\logm>10.8$ and $z<0.4$ in K21, due to their faster size growth at the massive end.

Intriguingly, the hosts of SDSS quasar have some overlap with the location of compact SFGs as shown in Figure 2 of \cite{Barro2017} on the \sigmae--\m~plane, which are most likely to be the progenitors of compact quiescent galaxies. We thus reach a similar conclusion as in Section \ref{subsec:size} that quasar hosts are more compact than the general inactive star-forming populations. The compact SFGs in \cite{Barro2017} overlap with quiescent galaxies on the $\Sigma_1-\m$ plane. Unfortunately, due to the limited resolution of HSC, we are not able to resolve $\Sigma_1$ for our quasars. It is also clear that the \sigmae--\m~relations for quasar hosts resemble that of SFGs, i.e., the surface-mass density increases with stellar mass. 
Such a trend is very different from massive quiescent galaxies, for which the \sigmae~decreases significantly beyond the pivot mass. This is likely due to dry minor mergers, which are more efficient in boosting the size growth while contributing less to the stellar mass assembly ($\Delta {\rm log}\,\re \propto 2\Delta\,\logm$), as well as adiabatic expansion of quiescent galaxies and the addition of newly quenched large galaxies \citep{Naab2009, Carollo2013, vanderwel2014, Barro2017}. The very different \sigmae--\m~relation between quasar hosts and massive quiescent galaxies indicates a distinct growth pattern for the two populations.

On the other hand, the increment in \sigmae~found for quasar hosts and SFGs is naturally expected for galaxies that have a shallow \re--\m~relation ($\alpha<0.5$), as 
\begin{equation}
    \Delta {\rm log\,}\re = \alpha \cdot \Delta\,\logm,
\end{equation}
thus
\begin{equation}
    \Delta {\rm log\,}\sigmae = \Delta\,\logm - 2\,\Delta\log\,\re 
\end{equation}
can be written as 
\begin{equation}
    \Delta {\rm log\,}\sigmae = (1-2\alpha) \cdot \Delta\,\logm,
\end{equation}
which exhibits a positive gradient at $\alpha<0.5$. Therefore, galaxies will become more compact as they continue to form stars as long as they evolve along a shallow \re--\m~track \citep{vandokkum2015, Barro2017}. The lack of significant evolution in the slope and normalization of the \re--\m~relations for SFGs (see Table 1 in M19) and SDSS quasars (Table \ref{table:fit}) ensure this. A shallow \re--\m~relation (or the corresponding \sigmae--\m~relation), however, is likely to be a result of several complicated mechanisms working together, including continuous gas accretion, gas compaction and subsequent star-formation, multiple mergers, effects of supernovae and AGN feedback, and further regulation from their dark matter halos \citep{Shen2003, vandokkum2015, Zolotov2015, Walters2021}. The effect of building the central mass concentration through these processes must be more frequent or efficient in quasar systems in order to explain the elevated \sigmae~for quasar host galaxies relative to inactive SFGs.
The implications of these results will be further discussed in the following section.

\section{Discussion}
\label{sec:discussion}
\subsection{On the Compact Nature of Quasar Host Galaxies}
The energy output from accreting black holes has long been speculated to play an important role in shaping the evolution of galaxy structural \citep[e.g.,][]{vanderVlugt2019, Zinger2020} and star formation \citep[e.g.,][]{Fabian2012, Dubois2013, Beckmann2017} properties, but whether the triggering of quasar activity preferentially occurs in galaxies with particular structural properties remains unclear. With a significant sample of luminous type-1 quasars from the SDSS DR14 quasar catalog, we have demonstrated that quasar hosts in general have low \ss~indices, suggesting the presence of disks (Figure \ref{fig:params}).  We have also found that their sizes lie between those of inactive SFGs and QGs at a given stellar mass (Figure \ref{fig:mass_size}).  Considering the average dependence of the bulge-to-total ratio on the \sid~from a single \ss~fit \citep[e.g.,][]{Dimauro2018, Ding2020}, the overall shape of the \sid~distribution may suggest that these quasars are still building up their bulges or are pseudobulges \citep[e.g.,][]{Yesuf2020b}, as many of them have \ss~indices lying between that of pure disks ($n=1$) and pure bulges ($n=4$) \citep[e.g.,][]{Gabor2009}.  Interestingly, this is in contrast to the fact that the majority of the massive SMBHs in the local universe are at the centers of ellipticals and classical bulges \cite[e.g.,][]{Gadotti2009, Kormendy2013}. 
At first glance, this could be consistent with a scenario in which quasar host galaxies are in a transition phase in which strong AGN feedback is responsible for turning blue, disk-dominated galaxies into red, compact ellipticals. However, the relatively blue colors for our quasars (Figure \ref{fig:uvj}) as well as recent findings that luminous AGNs are typically hosted by star-forming hosts \citep[e.g.,][]{Bernhard2019, Schulze2019, Xie2021} suggest that it is not very likely that quasar host galaxies have had their star formation quenched. 

It is more likely that our quasar hosts are gas-rich systems \citep[e.g.,][]{Vito2014, Shangguan2019, Shangguan2020, Yesuf2020, Koss2020}. They might be undergoing central mass build-up through active nuclear star formation \citep[e.g.,][]{Rujopakarn2018}. The observed compact sizes and elevated surface mass densities of quasar hosts relative to the general star-forming populations could occur if quasars are preferentially fueled in galaxies with prior concentrated gas reservoirs, whereby increasing the infall probability of a gas cloud. It may also caused by a dynamical compaction process described in theoretical models \citep[e.g.,][]{Dekel2014}, which has been widely adopted to explain the formation of compact galaxies at $z\sim1.5-3$ \citep[e.g.,][]{Barro2013, Barro2017}.

In the compaction model, substantial amounts of cold, perturbed streams undergo dissipative collapse and flow toward the nuclear region when they lose angular momentum through mergers, counter-rotation, and/or violent disk instabilities \citep{Hopkins2006, Dekel2014, Bournaud2011, Zolotov2015}. The inflowing gas thus leads to a centrally confined starburst and is also likely to fuel SMBH accretion \citep[e.g.,][]{Bournaud2011, Rujopakarn2018, Lapiner2020}. A spatial redistribution of disk stars to the central region could also help to build a bulge. This picture is consistent with our results and previous findings from both observations \citep[e.g.,][]{Rangel2014, Kocevski2017, Jahnke2009,Ni2019, Silverman2019} and large-volume cosmological simulations \citep[e.g.,][]{Habouzit2019} that AGNs are most commonly detected in compact SFGs. 
Recent far-infrared and submillimeter observations taken by the Herschel Space Observatory and the Atacama Large Millimeter/Submillimeter Array have also revealed the compact nature of the gas and dust content in AGN hosts, further suggesting that the centrally concentrated cold gas may be providing fuel for both black hole accretion and nuclear star formation \citep[e.g,][]{Lutz2018, Puglisi2019, Chang2020, Stacey2020}. 

Although the efficiency and frequency of occurrence of a compaction process are likely to diminish with time due to the decreasing cold gas reservoir and/or merger rate, the more compact nature of quasar hosts found here indicates that the concentrated gas distribution may still be an important condition in triggering bright quasar episodes at $z<1$. These quasars are probably at an early stage of the compaction phase and have not yet evolved to compact quiescent galaxies \citep{Barro2013, Barro2014} as their colors are blue. In this regard, it may be that our quasars are signaling the formation of compact galaxies in the less gas-rich, low-redshift universe, as the same process also ignites quasar activity. Once the core-building event is completed, star formation in galaxies may weaken because of gas consumption, AGN feedback or morphological quenching \citep{Martig2009, Tacchella2015, Wu2020}.  Quasar hosts may thus become quiescent with an increased central mass concentration, which is consistent with our result that the quiescent hosts of quasars are smaller in size than star-forming ones (Figure \ref{fig:mass_size}).  Considering that the quiescence of a galaxy appears to be strongly connected to its central mass density \citep{Kauffmann2003, Fang2013, Whitaker2017}, our star-forming quasars may be right on their way to quench, as their surface mass densities are reaching those of quiescent galaxies (Figure \ref{fig:sigma_m}).

It is currently not clear whether the compaction, if present, is triggered by internal processes or is tidally induced through mergers. If driven by mergers, the high disk fraction among our quasar hosts would prefer minor mergers instead of major mergers as the disk structure is likely to be destroyed in a violent major merger event \citep{Hopkins2009disk}. On the other hand, the large dispersion of the size--mass relation suggests that quasar activity can be widely triggered in a variety of galaxy populations and processes.  The gas inflows driven by secular processes through, e.g., bars, nuclear spirals in extended SFGs or QGs may also lead to active black hole accretion, obviating the need for a violent compaction process. The low \ss~indices of quasar hosts may highlight the importance of disk structures in transforming angular momentum of the inflowing gas, as the inflow intensity is expected to weaken as the disk vanishes \citep[e.g.,][]{Zolotov2015}. 
Detailed investigations of host galaxy substructures (e.g., nuclear spirals, bars, disturbed features) are thus required to further understand how black holes obtain their fuel. This could be achieved by examining the residual images of low-redshift objects after subtracting the quasar and the smooth galaxy component (see Figure \ref{fig:example}). 
Disk-like structures are clearly seen in $\sim 60\%$ of the $z<0.4$ sources in our sample, consistent with the distribution of \ss~indices. We plan to investigate this topic further in future work. 

As an alternative to the compaction model, it could be that the positive AGN feedback is at work whereby stars are forming in the compressed interstellar medium or AGN-driven outflow. If the newly formed stars are gravitationally bound to the central region, they will contribute to the formation of bulges \citep{Ishibashi2014, Maiolino2017, Gallagher2019}. If the outflow scenario is driving the elevated central mass concentration and blue colors for quasar hosts,  we would expect to see larger sizes for quasars with higher luminosities. This is because the SFR within outflows is observed to be positively correlated with the outflow velocity \citep{Gallagher2019}, which in turn is proportional to AGN luminosity \citep{Fiore2017}. Therefore, luminous quasars will tend to deposit newly formed stars on larger scales.  While, we see no trend between size and quasar luminosity and Eddington ratio (Figure \ref{fig:size_bin}), at least at lower redshifts where we can robustly constrain the properties of quasars and their hosts, this mechanism may still have played a role. 

It is also possible that quasar host galaxies with the most massive black holes have grown earlier to accumulate their black hole masses, thus having an old bulge and a dense central region with systematically smaller size. 
At the same time, they may be still building their disks, making the hosts blue. This scenario has been described in \cite{Lilly2016} where an inside-out disk growth is able to reproduce the observed quiescence fraction on galaxy morphology. The existence of both a core-building compaction process which leads to recent star formation in galaxy centers and AGN activity, and an inside-out pattern of galaxy growth resulting in older cores and younger galaxy outskirts has been reported by \cite{Woo2019}.  However, the inside-out scenario appears to be inconsistent with the fact that the observed colors (e.g., $g-i$) of our quasar hosts in the central regions (e.g., $\sim$2 kpc) also tend to be blue, although we caution that such regions are unresolved in our images and a recent central burst in an old bulge could also make their colors blue. Spatially resolved investigations of the stellar ages of our quasar host galaxies are required to robustly distinguish between the two scenarios.

\subsection{Delay or No Quasar Feedback?}
From a theoretical perspective, AGN feedback is crucial in quenching star formation and reproducing the observed galaxy size and stellar density evolution. In models that do not include energy from AGNs which can heat or vacate the ISM of the galaxy, the star formation does not cease and the central region of the galaxies become overly blue, compact and massive \citep[e.g.,][]{Choi2018, Habouzit2019, vanderVlugt2019}.
Considering that the majority of quasars are hosted by SFGs (see Section \ref{subsec:sf_prop}), we would expect quasar hosts to be larger than SFGs without AGN if the AGN feedback-induced star formation and/or mass loss caused a significant expansion of the galaxy. However, we see the opposite: quasar hosts are more compact on average than SFGs. In addition, the distribution of sizes of high luminosity/Eddington ratio quasars are indistinguishable from those of low luminosity/Eddington ratio (Figure \ref{fig:size_bin}), conflicting with our expectation that the feedback from more powerful quasars would have a larger impact on galaxy size. As a result, AGN feedback is unlikely to drive the size growth of the galaxies as predicted in some theoretical models \cite[e.g.,][]{Fan2008, Ishibashi2013}. Instead, it appears that our quasars are rather a byproduct of a galaxy evolution phase possibly triggered by dynamical compaction, signaling the process of transforming stellar mass from the low-\sid~disk component to the centrally concentrated bulge component through nuclear star formation or a redistribution of stars \citep[e.g.,][]{Jahnke2009, Silverman2019, Ding2020}. 

Nonetheless, the effects of AGN feedback on galaxy size growth may not have fully appeared yet. The typical timescale for an individual episode of AGN activity is short ($\sim 10^5 - 10^6$ yrs; \citealt{Keel2012, Keel2015, Schawinski2015}). The timescale for secular expansion of the galaxy via AGN feedback is likely to be rather longer (e.g., $\sim 2$~Gyrs in \citealt{Fan2008}). Cosmological simulations \citep[e.g.,][]{Yuan2018, Choi2018, Zinger2020, Terrazas2020} have shown that the galaxies grow in size when black holes switch to a radiatively inefficient hot accretion mode, during which the cumulative energy injected by AGN feedback exceeds the gravitational binding energy of gas within galaxies and start to play a role in regulating galaxy structure and star formation. Therefore, although the energy released by the quasar phase may not be sufficient and there may not have been enough time to initiate strong galaxy size growth when both black hole accretion and star formation are still in an active phase, it is possible that feedback effects will eventually emerge after some time delay \citep[e.g.,][]{Yesuf2014, Woo2017}.

Furthermore, the lack of direct observational evidence for feedback-induced size growth may be because AGNs are elevating galaxy sizes indirecly. Specifically, from a theoretical perspective, AGN feedback can affect the stellar density profile in two major ways. First, AGN-driven outflows can induce fluctuations in the gravitational potential of the central region of the galaxy, rearranging the stellar component to a new equilibrium with a puffed-up core \citep[e.g.,][]{Fan2008}. Second, the large amount of feedback energy and momentum being injected into the ISM can reduce the cold gas fraction, thereby increasing the frequency of dry minor mergers which have been argued to be a principal mechanism for massive galaxies like our quasar hosts to grow further \citep[e.g.,][]{Naab2009, Sanjuan2012, Oogi2013, Faisst2017, Damjanov2019}. The latter mechanism has been argued to play a major role in regulating galaxy size evolution \citep[e.g.,][]{Dubois2013, Choi2018}.  Since merger-induced size growth is most effective for gas-poor galaxies (i.e., dry mergers), and our quasars which are still actively forming stars have not yet evolved to such a stage, the resulting growth 
lies in the galaxies' future. Combining structural analyses for various AGN populations in different stages of evolution, such as heavily obscured AGNs which may represent an early fast growth phase of SMBHs \citep{Kocevski2015, Li2020}, post-starburst quasars which connect quasars to recently quenched galaxies \citep[e.g.,][]{Cales2015, Matsuoka2015}, radio AGNs which represent the low state of black hole accretion \citep[e.g.,][]{Best2005}, and galaxies showing past AGN activity like those with extended emission line regions \citep[e.g,][]{Keel2012, Ichikawa2019a, Ichikawa2019b} may complete the evolutionary path and shed light on how AGN activity may affect galaxy evolution.

\section{Summary}
\label{sec:summary}

We present a structural analysis of a parent sample of $\sim5000$ quasar host galaxies at $0.2<z<1.0$ drawn from the SDSS DR14 quasar catalog, using optical imaging in five bands ($grizy$) from the HSC-SSP survey which uniquely combines wide survey area, deep image depth and superb spatial resolution. A two-dimensional \ss~profile and PSF model are used to decompose the images into an underlying host galaxy and the point source component, allowing us to measure the structural parameters and photometric properties of quasar host galaxies. Detailed image simulations are performed to calibrate and assess our measurements using both real and model galaxies. Our main conclusions are as follows:
\begin{enumerate}

\item The host galaxy contributes significantly to the total light measured in the HSC \i-band, with a mean host-to-total flux ratio decreasing from $\sim70\%$ at $z=0.3$ to $\sim30\%$ at $z=0.9$ (Section \ref{subsec:host_ratio}).

\item Quasar hosts are preferentially biased toward face-on systems, suggesting that galactic-scale dust can contribute to obscuring the broad line region (Section \ref{subsec:ellp}).

\item The overall \sid~distribution (a peak value of $\sim 1.0$ and a median value of $\sim 2.0$) is consistent with quasars residing in galaxies with disk-like light profiles over the full redshift range considered (Section \ref{subsec:sersic_index}).

\item Quasar host galaxies preferentially lie in a region on the rest-frame $U-V$ vs. $M_*$ diagram that is bluer than the red sequence, and more massive than the blue cloud. The fractions of star-forming hosts are estimated to be $\sim70\%$ over the full redshift range considered. Therefore, we find that SDSS quasars are preferentially hosted by massive star-forming galaxies (Section \ref{subsec:sf_prop}).

\item While quasar hosts have a wide range of sizes at a given \m, their average values follow a positive linear relation thus exhibit a size -- mass relation as seen in the inactive galaxy population (Section~\ref{subsec:size}). 

\item The average size -- mass relation of quasar hosts is intermediate between those of inactive SFGs and QGs of similar stellar mass. This is not due to a mixture of host types since quasars with SF hosts also show smaller sizes than the inactive SF galaxies (Section~\ref{subsec:size}).

\item The average sizes of high luminosity (or \edd~and \mbh) quasars are indistinguishable from that of less luminous quasars at a given stellar mass (Section~\ref{subsec:size}). 

\item Taken together, the results outlined above demonstrate that quasars prefer to lie in compact SFGs. This is further illustrated by their surface mass density that exhibits intermediate values between the inactive star-forming and quiescent galaxy populations, with the mean increases with stellar mass (Section \ref{subsec:density}).

\end{enumerate}

These findings are consistent with a scenario in which quasar activity is preferentially triggered in galaxies with centrally concentrated gas distributions. This can occur in galaxies with prior concentrated gas reservoirs, or could be driven by a dynamical compaction process through, e.g., disk instabilities and minor mergers, during which the inflowing gas and stars from the disk concurrently build up the bulge and ignite active black hole accretion. On the other hand, our observed size--mass relationship of quasar host galaxies and the fact that there are no significant size differences between luminous and less luminous quasars appear to be inconsistent with the hypothesis that AGN-driven mass loss and/or positive feedback induced star formation on larger scales can cause a significant expansion in galaxy size.

However, the distributions of size, compactness and \sid~are broad, indicating that a compaction phase may not be all that is required. Other processes such as continuous/stochastic accretion through, e.g., bars, nuclear spirals may also drive gas inward and trigger an AGN. The elevated central mass concentration and blue colors could also be a result of positive AGN feedback, if the newly formed stars driven by AGN outflows are gravitationally bound to the galaxy centers. To disentangle the various AGN triggering mechanisms and reveal the role played by AGN feedback in regulating galaxy structures, we need to carry out detailed studies of the substructures of AGN host galaxies and their large-scale environment while comparing them to those of inactive galaxies. This could be achieved by investigating the residual images after subtracting the point source and the smooth galaxy components (see Figure \ref{fig:example}). Observations of molecular gas and star formation in the nuclear region for a large sample of AGNs with a range of physical properties and evolutionary stages will also be required to better understand how the central gas concentration is established and is connected to SMBH activity. 

\acknowledgments 
We thank the anonymous referee for helpful comments that helped us to improve the quality of the paper.
J.Y.L thanks Naoki Yasuda for his help in analyzing HSC imaging data.
J.Y.L. is supported by the China Scholarship Council. J.D.S. is supported by the JSPS KAKENHI Grant Number JP18H01251, and the World Premier International Research Center Initiative (WPI Initiative), MEXT, Japan. J.Y.L. and Y.Q.X. acknowledge support from the National Natural Science Foundation of China (NSFC-12025303, 11890693, 11421303), the CAS Frontier Science Key Research Program (QYZDJ-SSW-SLH006), the K.C. Wong Education Foundation and the Chinese Space Station Telescope (CSST) Project.

The Hyper Suprime-Cam (HSC) collaboration includes the astronomical communities of Japan and Taiwan, and Princeton University. The HSC instrumentation and software were developed by the National Astronomical Observatory of Japan (NAOJ), the Kavli Institute for the Physics and Mathematics of the Universe (Kavli IPMU), the University of Tokyo, the High Energy Accelerator Research Organization (KEK), the Academia Sinica Institute for Astronomy and Astrophysics in Taiwan (ASIAA), and Princeton University. Funding was contributed by the FIRST program from Japanese Cabinet Office, the Ministry of Education, Culture, Sports, Science and Technology (MEXT), the Japan Society for the Promotion of Science (JSPS), Japan Science and Technology Agency (JST), the Toray Science Foundation, NAOJ, Kavli IPMU, KEK, ASIAA, and Princeton University. 
This paper makes use of software developed for the Large Synoptic Survey Telescope. We thank the LSST Project for making their code available as free software at  http://dm.lsst.org
The Pan-STARRS1 Surveys (PS1) have been made possible through contributions of the Institute for Astronomy, the University of Hawaii, the Pan-STARRS Project Office, the Max-Planck Society and its participating institutes, the Max Planck Institute for Astronomy, Heidelberg and the Max Planck Institute for Extraterrestrial Physics, Garching, The Johns Hopkins University, Durham University, the University of Edinburgh, Queen’s University Belfast, the Harvard-Smithsonian Center for Astrophysics, the Las Cumbres Observatory Global Telescope Network Incorporated, the National Central University of Taiwan, the Space Telescope Science Institute, the National Aeronautics and Space Administration under Grant No. NNX08AR22G issued through the Planetary Science Division of the NASA Science Mission Directorate, the National Science Foundation under Grant No. AST-1238877, the University of Maryland, and Eotvos Lorand University (ELTE) and the Los Alamos National Laboratory.
Based [in part] on data collected at the Subaru Telescope and retrieved from the HSC data archive system, which is operated by Subaru Telescope and Astronomy Data Center at National Astronomical Observatory of Japan.
This work is based on observations taken by the 3D-HST Treasury Program (GO 12177 and 12328) with the NASA/ESA HST, which is operated by the Association of Universities for Reserach in Astronomy, Inc., under NASA contract NAS5-26555.

\software{CIGALE \citep{Boquien2019}, emcee \citep{Mackey2013}, hscpipe (v6.7; \citealt{Bosch2018}), Photutils \citep{Bradley2019}, Lenstronomy \citep{Birrer2015, Birrer2018}}

\appendix
\section{Correcting for Internal Color Gradients}
\label{appendix:a}

Due to the presence of internal color gradients in galaxies, the galaxy size depends on observed wavelength \citep[e.g.,][]{vanderwel2014, Vulcani2014}. Following \cite{vanderwel2014}, we correct the \i-band size measurements to a common rest-frame wavelength of 5000~\ang~to account for this effect. The corrected size is parameterized as
\begin{equation}
r_{e,c} = r_{e,i} \Big(\frac{1+z}{1+z_p}\Big)^{\frac{\Delta\,{\rm log}\,r_e}{\Delta {\rm log}\,\lambda}},
\label{eq:size}
\end{equation}
where $r_{e,c}$ denotes the size being calibrated to 5000~\ang, $r_{e,i}$ represents the \i-band half-light radius, $\Delta\,{\rm log}\,r_e / \Delta {\rm log}\,\lambda$ is the color gradient, and $z_p$ is the ``pivot redshift'' at which the rest-frame effective wavelength of a given filter is equal to 5000~\ang~(for our choice of the HSC \i-band, $z_p$ is 0.55). The color gradient for quasars hosted by SFGs is formulated as 
\begin{equation}
\frac{\Delta\,{\rm log}\,r_e}{\Delta {\rm log}\,\lambda} = -0.35 + 0.12z - 0.25\,{\rm log}\,\Big(\frac{M_*}{10^{10}\,M_\odot}\Big),
\end{equation}
while for quasars hosted by QGs we utilize $\Delta\,{\rm log}\,r_e / \Delta {\rm log}\,\lambda = -0.25$ with no dependence on redshift or stellar mass. The star-forming and quiescent hosts are determined based on the rest-frame $U-V$ vs. \m~diagram (see Sections \ref{subsec:sed} and \ref{subsec:sf_prop}). 

To verify that the wavelength-dependent size calibration derived from HST imaging data for CANDELS galaxies \citep{vanderwel2014} is appropriate for our HSC quasars, we refit \r-band images at $0.2<z<0.3$ and \z-band images at $0.7<z<0.8$ by allowing the \ss~parameters to vary freely. At these redshift ranges, the \r- and \z-band images approximately probe rest-frame 5000 \ang, thereby a direct comparison between \r- and \z-band sizes ($r_{e,r}$ and $r_{e,z}$, respectively) with $r_{e,c}$ can be made.

At $0.2<z<0.3$, the median values of $r_{e,i}$, $r_{e,c}$ and $r_{e,r}$ are 4.2~kpc, 4.6~kpc and 4.6~kpc, respectively; while for $0.7<z<0.8$, the median values of $r_{e,i}$, $r_{e,c}$ and $r_{e,z}$ are 4.9~kpc, 4.6~kpc and 4.6~kpc, respectively. The larger (smaller) size measured from the \r-band (\z-band) images than the \i-band result is consistent with a picture that bluer (redder) band traces younger (older) stellar populations thereby having more extended (concentrated) light distributions. The good consistency between $r_{e,c}$ with $r_{e,r}$ and $r_{e,z}$  suggests that the wavelength dependence of \re~derived in \cite{vanderwel2014} is applicable to our quasar 
host galaxies.

\section{Recovery of S$\rm \acute{e}$rsic index, ellipticity and host-galaxy magnitude using image simulations}
\label{appendix:b}

\begin{figure*}
\centering
\includegraphics[width=\linewidth]{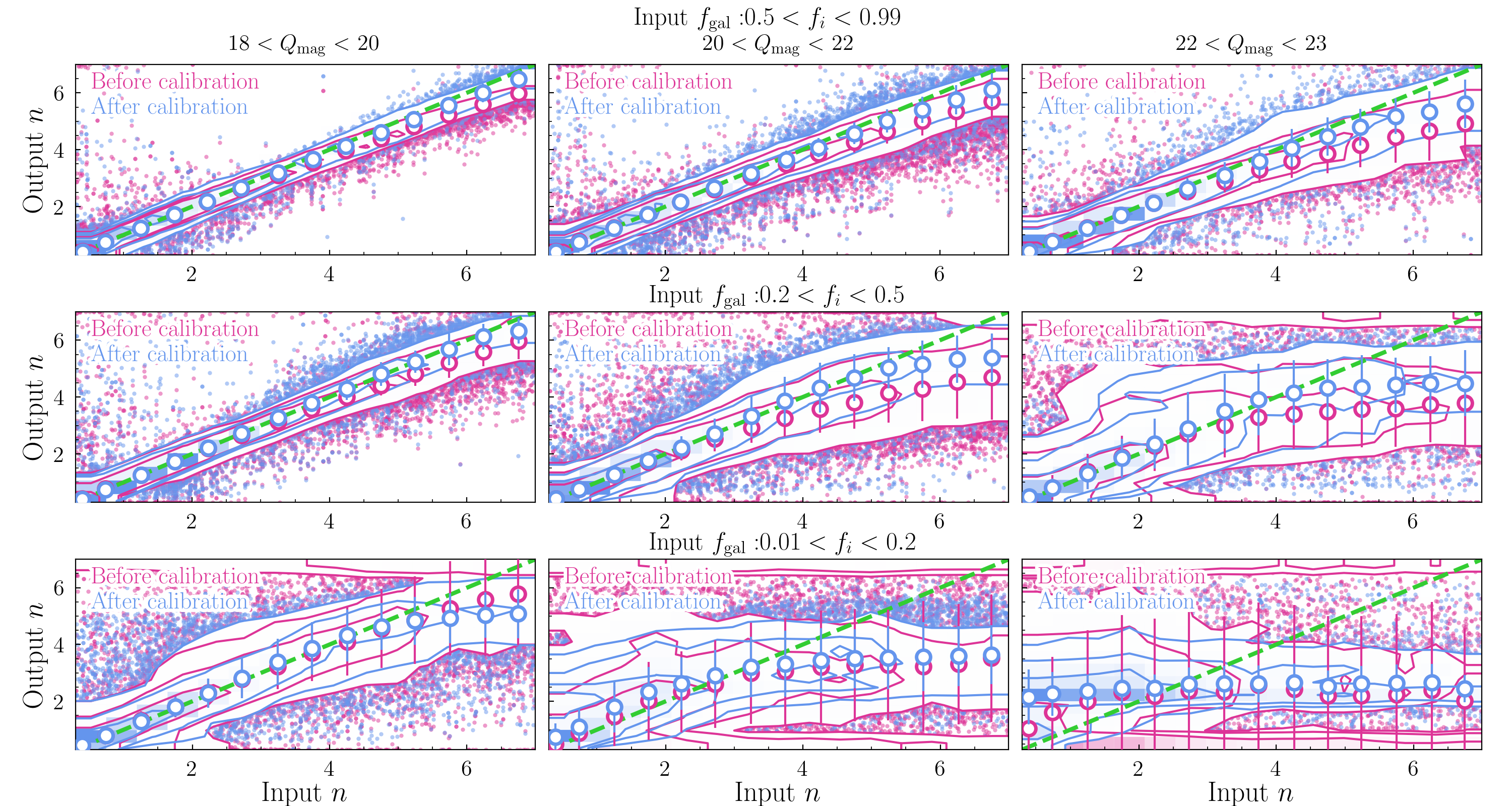}
\caption{Similar to Figure \ref{fig:input_output} but for the \sid~in the \i-band.}
\label{fig:input_output_n}
\end{figure*}

Using simulations described in Section~\ref{subsec:simu}, we compare the input \ss~indices to the measured values before and after calibration (Figure~\ref{fig:input_output_n}). The scatter of the fitted values around the input values are large for faint hosts ($\fr<0.5$). The calibration procedure can on average recover the true \sid, but tends to underestimate the \sid~for faint hosts with $n>4$.

\begin{figure*}
\centering
\includegraphics[width=\linewidth]{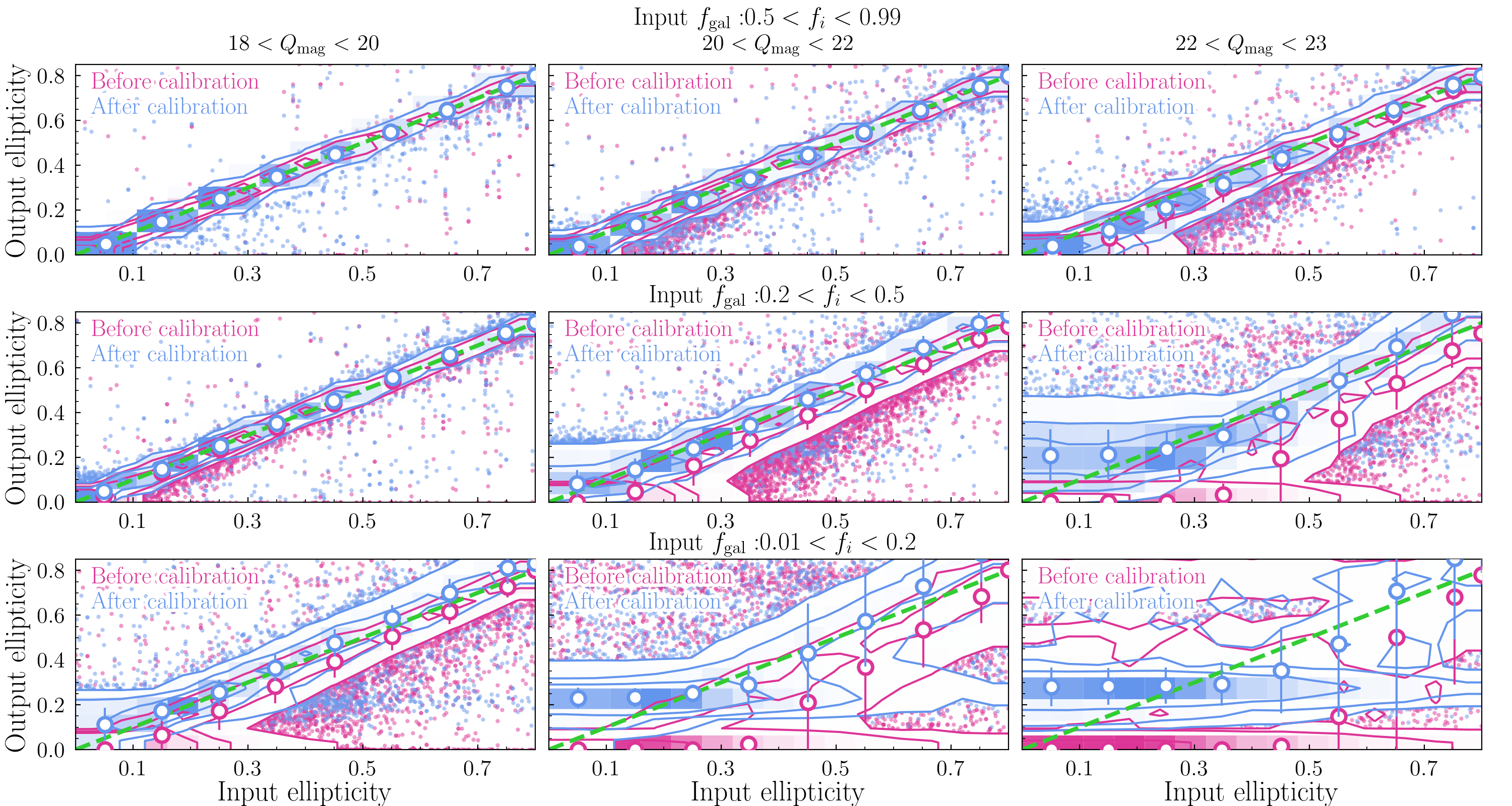}
\caption{Similar to Figure \ref{fig:input_output} but for the galaxy ellipticity in the \i-band.}
\label{fig:input_output_e}
\end{figure*}

Figure \ref{fig:input_output_e} compares the fitted and calibrated galaxy ellipticity to the input values. The ellipticity tends to be significantly underestimated when the host galaxy is fainter than $\sim22$~mag. However, our calibration procedure is able to effectively recover the intrinsic ellipticity, thus elliminating the biases when examining the orientation of quasar host galaxies (Section \ref{subsec:ellp}).

\begin{figure*}
\centering
\includegraphics[width=\linewidth]{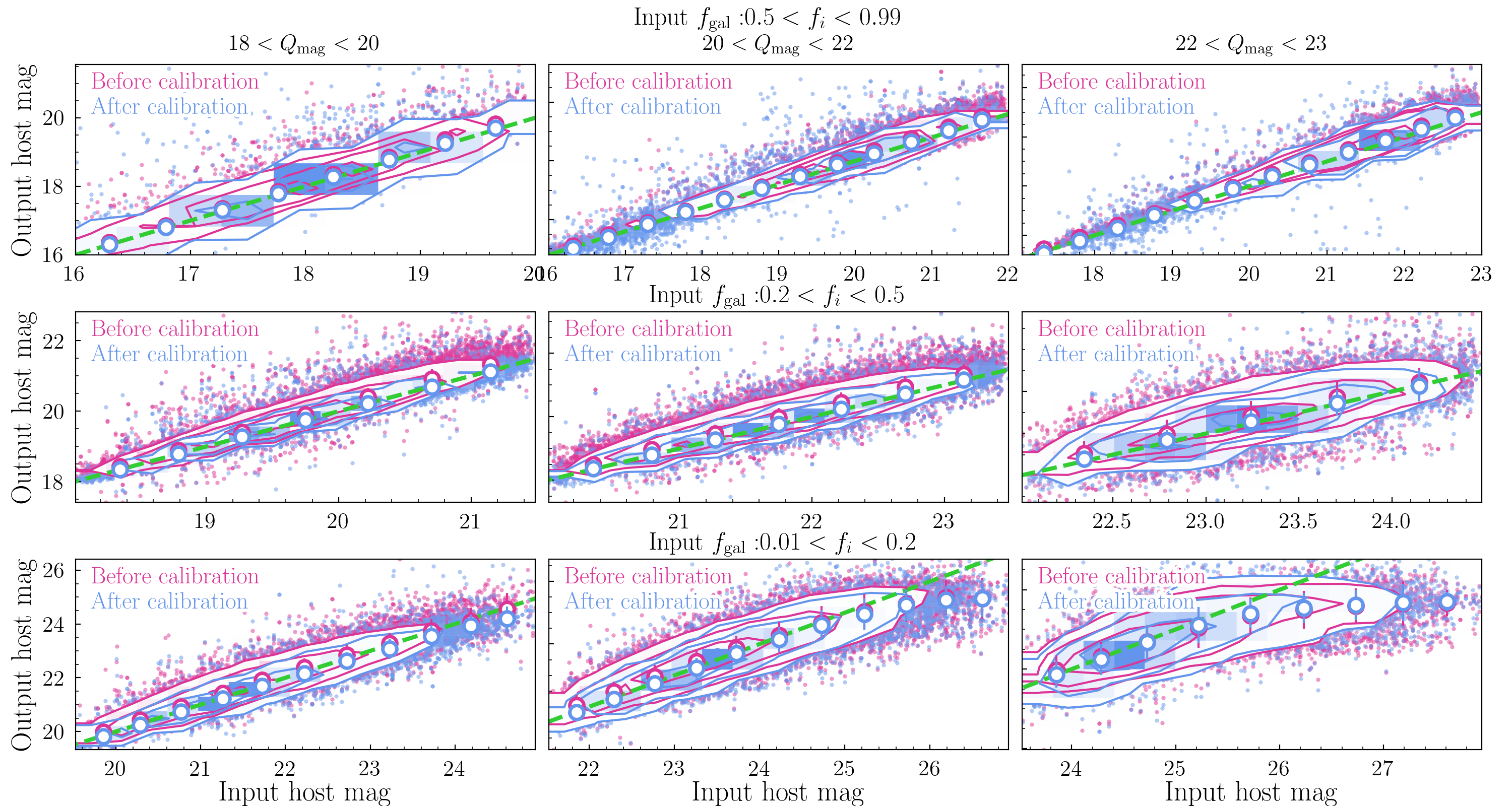}
\caption{Similar to Figure \ref{fig:input_output} but for the decomposed galaxy magnitude in the \i-band.}
\label{fig:input_output_hmag}
\end{figure*}

The comparison between the input \i-band host galaxy magnitudes with the fitted and calibrated values are plotted in Figure \ref{fig:input_output_hmag}. We find that the galaxy magnitudes can be recovered well in nearly all cases with the exception of those with very faint hosts (i.e., the bottom right panel), demonstrating that the total galaxy flux is a robust measurement from our decomposition, at least for the \i-band with the highest image quality.

 \newcommand{\noop}[1]{}

\begin{table*}
\renewcommand{\arraystretch}{1.2}
\small
\caption{Catalog of 2D structural measurements and SED fitting results}
\small
\centering
 \begin{tabular}{ll} 
 \hline
 \hline
 Column Name & Description \\
 \hline
 ID & Unique identifier from the SDSS DR14 quasar catalog \\ 
 
 RA & right ascension in degree \\
 
 Dec & declination in degree\\
 
$z$ & redshift from the SDSS DR14 quasar catalog\\
 
host\_[$grizy$]mag & decomposed host-galaxy magnitude in $grizy$ bands\\

ps\_[$grizy$]mag & decomposed point-source magnitude in $grizy$ bands\\

$f_{\rm gal}\_[grizy]$ & host-to-total flux ratio in $grizy$ bands\\


$r_{\rm e}$ & effective radius in arcsec and measured in the \i-band (no correction applied)\\

$r_{\rm e}^c$ & effective radius in arcsec and measured in the \i-band (with correction applied; see text for details)\\

$n_s$ & Sersic index in the \i-band (no correction applied)\\

$n_s^c$ & Sersic index in the \i-band (with correction applied)\\

$n_{flag}$ & flag descriptive of whether the \sid~parameter was fixed (0=no, 1=yes)\\

ellp & ellipticity in the \i-band\\

PA & position angle in radian in the \i-band (East of North)\\

$\logm$ & logarithm of stellar mass from CIGALE\\

color & rest-frame $U-V$ color from CIGALE\\

type & "SF" for star-forming and "QS" for quiescent host galaxies\\

final & whether the source is in our final sample as described in Section \ref{sec:final_sample}\\

\hline
\end{tabular}
\label{table:catalog}
\end{table*}

\begin{table*}
\renewcommand{\arraystretch}{1.2}
\caption{SDSS/HSC quasar sample statistics (see Section~\ref{sec:final_sample} for details).}
\centering
\begin{tabular}{lcccc}
\hline
\hline
Criterion & $0.2<z<0.4$ & $0.4<z<0.6$ & $0.6<z<0.8$ & Total\\
\hline
Full sample & 347 & 873 & 1756 & 2976\\
\hline
$M_*$ cut (lower) & 5 & 58 & 339 & 402\\
$M_*$ cut (upper) & 0 & 5 & 29 & 34\\
\i-band detected & 2 & 6 & 13 & 21\\
$\chi^2_{\rm SED}$ cut & 3 & 5 & 26 & 34\\
$\Delta \rm SB$ cut & 22 & 14 & 4 & 40\\
$\omega$ cut & 0 & 0 & 10 & 10\\
$N_{\rm simu}$ cut & 9 & 2 & 0 & 11\\
\hline
Total number excluded & 41 & 90 & 421 & 552\\
\hline
Final sample & 306 & 783 & 1335 & 2424\\
\hline
\end{tabular}
\label{table:exclude}
\end{table*}

\begin{table*}
\centering
\renewcommand{\arraystretch}{1.4}
\caption{Best-fit parameters of single-power-law fit in the form of Equation \ref{eq:pwl} to the size--mass relation for all quasars, star-forming quasars and quiescent quasars down to the stellar mass cut. Column (1) samples used to perform analytic fits, (2) $z_{\rm med}$ is the median redshift, (3) ${\rm log}\,(M_{\rm cut}/M_\odot)$ is the stellar mass cut, (4, 5, 6) $A$, $\alpha$ and $\sigma_{{\rm log}\,r}$ are the intercept, slope and intrinsic scatter of the \re--\m~relation. The uncertainties on parameters which are less than 0.01 are indicated as 0.}
\begin{tabular}{cccccc}
\hline
\hline
Sample & $z_{\rm med}$ & ${\rm log}\,(M_{\rm cut}/M_\odot)$ & $A$ & $\alpha$ & $\sigma_{{\rm log}\,r}$\\
(1) & (2) & (3) & (4) & (5) & (6)\\
\hline
 &  0.3 & 9.3  & $5.6_{-0.1}^{+0.1}$ & $0.37_{-0.03}^{+0.03}$ & $0.17_{-0.01}^{+0.01}$ \\
All  &  0.5 & 9.8 & $4.9_{-0.1}^{+0.1}$ & $0.33_{-0.02}^{+0.02}$ & $0.17_{-0.01}^{+0.01}$ \\
 &  0.7 & 10.3 & $4.8_{-0.1}^{+0.1}$ & $0.29_{-0.02}^{+0.02}$ & $0.19_{-0.00}^{+0.00}$ \\
 \hline
  &  0.3 & 9.3 & $6.2_{-0.3}^{+0.2}$ & $0.34_{-0.04}^{+0.04}$ & $0.14_{-0.01}^{+0.01}$ \\
Star-forming  &  0.5 & 9.8 & $5.3_{-0.2}^{+0.1}$ & $0.33_{-0.03}^{+0.03}$ & $0.18_{-0.01}^{+0.01}$ \\
 &  0.7 & 10.3 & $5.1_{-0.1}^{+0.2}$ & $0.26_{-0.03}^{+0.03}$ & $0.18_{-0.01}^{+0.01}$ \\
 \hline
  &  0.3 & 9.3 & $4.8_{-0.2}^{+0.2}$ & $0.54_{-0.06}^{+0.06}$ & $0.16_{-0.01}^{+0.01}$ \\
Quiescent  &  0.5 & 9.8 & $4.5_{-0.2}^{+0.2}$ & $0.39_{-0.04}^{+0.04}$ & $0.15_{-0.01}^{+0.01}$ \\
 &  0.7 & 10.3 & $4.2_{-0.1}^{+0.1}$ & $0.37_{-0.05}^{+0.05}$ & $0.20_{-0.01}^{+0.01}$ \\
\hline
\end{tabular}
\label{table:fit}
\end{table*}

\end{document}